\documentclass[floatfix,aps,twocolumn,showpacs,nofootinbib,superscriptaddress]{revtex4} 

\usepackage{graphicx}
\usepackage{subfigure}
\usepackage{epsfig} 
\usepackage{amsmath} 
\usepackage{amsfonts}
\usepackage{amssymb} 
\usepackage{hyperref}
\usepackage{dsfont}

\newcommand{\mean}[1]{\langle #1 \rangle}
\DeclareMathOperator{\Tr}{Tr}
\renewcommand\Re{\operatorname{Re}}
\renewcommand\Im{\operatorname{Im}}

\begin{document}

\author{Nahuel Freitas} 
\affiliation{Departamento de F\'\i sica, FCEyN, UBA, Pabell\'on 1, 
Ciudad Universitaria, 1428 Buenos Aires, Argentina}
\affiliation{Instituto de F\'\i sica de Buenos Aires, UBA CONICET,  
Pabell\'on 1, Ciudad Universitaria, 1428 Buenos Aires, Argentina} 

\author{Juan Pablo Paz}
\affiliation{Departamento de F\'\i sica, FCEyN, UBA, Pabell\'on 1, 
Ciudad Universitaria, 1428 Buenos Aires, Argentina}
\affiliation{Instituto de F\'\i sica de Buenos Aires, UBA CONICET,  
Pabell\'on 1, Ciudad Universitaria, 1428 Buenos Aires, Argentina} 

\title{Fundamental limits for cooling of linear quantum refrigerators}

\date{\today}

\begin{abstract} 
We study the asymptotic dynamics of arbitrary linear quantum open 
systems which are periodically driven while coupled with generic  
bosonic reservoirs. We obtain exact results for the heat flowing 
into the network, which are valid beyond the usual weak coupling or 
Markovian approximations. We prove the validity of the dynamical 
third law of thermodynamics (Nernst unattainability principle), 
showing that the ultimate limit for cooling is imposed by a 
fundamental heating mechanism which becomes dominant at low 
temperatures: the non resonant creation of pairs of excitations 
in the reservoirs induced by the driving field. This 
quantum effect, which is missed in the weak coupling approximation, 
restores the unattainability principle whose validity was recently challenged. 
\end{abstract}

\maketitle 

{\it Introduction:} Is there a 
 fundamental limit
for cooling? The third law 
of thermodynamics
formulated by Nernst \cite{nernst1926} (the unattainability 
principle) establishes that it is not 
possible to reach zero temperature in 
finite time.  
But this is not a 
sacred principle and, in fact, its validity has 
been challenged 
\cite{kolavr2012, levy2012, cleuren2012}
on the basis of new results obtained in the 
emergent field of ``quantum thermodynamics'' 
\cite{kosloff2013, anders2015, brandao2015}. Challenging
macroscopic thermodynamical laws is 
not a luxury but a necessity 
when studying energy exchange 
in a regime dominated by quantum 
fluctuations. Using modern 
quantum technologies, 
systems evolving in 
these conditions (often 
out of equilibrium and 
far from the thermodynamical limit),
are now regularly 
built and manipulated
\cite{abah2012,an2015,pekola2015}. 
Thus, these technologies 
enabled a new generation
of machines, operating as engines or
refrigerators in the quantum domain. In fact, 
to understand  
devices such as the single ion engine\cite{abah2012},
the quantum absorption refrigerators\cite{levy2012refri}, 
the phononic coolers\cite{arrachea2012}, 
or the quantum thermal transistor\cite{joulain2016}, 
quantum thermodynamics is essential.

In this Letter we show that, contrary
to previous claims  \cite{kolavr2012, cleuren2012},  the 
unattainability principle is indeed valid
for a wide class of quantum 
refrigerators which includes the
most prominent cases mentioned above 
(phononic and ionic coolers, for example). 
Most importantly, we show  
that the ultimate limit for cooling  
is imposed by a fundamental 
quantum process, which is 
next-to-leading order in the 
coupling between the 
system and the reservoirs and 
becomes dominant
at sufficiently low temperatures. It consists of 
the non resonant creation of excitation pairs 
in the reservoirs, 
which is closely related with
the dynamical Casimir effect (DCE)\cite{wilson2011,benenti2015}. 
As this process is not of leading order 
in the coupling between 
the system and the reservoirs, its 
fundamental role was missed in all 
previous works that were based on a weak 
coupling approximation. On the basis
of those analysis, potential 
violations of the third
law were suggested \cite{kolavr2012, cleuren2012}. 
Instead, the unavoidable heating process 
we characterize enforces the third law. 

We present exact results for   
quantum refrigerators formed by 
networks of oscillators  
whose frequencies and
couplings are driven by external sources. 
Their Hamiltonian is 
$H_S(t)=(P^T M^{-1} P + X^T V(t) X)/2$ 
($X$ and $P$ are vectors whose components 
are position and momentum operators  
satisfying canonical commutation relations, 
being $M$ and $V(t)$ positive and symmetric matrices. The network is coupled 
with arbitrary bosonic reservoirs $E_\alpha$
and the total Hamiltonian is  
$H_T = H_S(t) + \sum_\alpha H_{E_\alpha} + \sum_\alpha H_{int,\alpha}$. Each reservoir $E_\alpha$ is
formed by independent oscillators whose 
positions  (momenta) are  
$q_{\alpha,j}$ ($\pi_{\alpha,j}$). Thus, 
$H_{E,\alpha} = \sum_{j}
 ({\pi_{\alpha,j}^2}/{2m}
+ {m\omega_{\alpha,j}^2} q_{\alpha,j}^2/2)$ 
and the interaction is described by the bilinear
Hamiltonian
$H_{int,\alpha} = \sum_{j,k} C_{\alpha,jk}\; X_j\:q_{\alpha,k}$  
(with coupling constants $C_{\alpha,jk}$). 

Work and heat, 
the two central thermodynamical concepts are
naturally defined once we notice that 
the expectation value of the energy of $S$,  
$\langle H_S\rangle$, changes
for two reasons. The variation   
induced by the external driving is 
associated with the work $W$, while  
the change induced by the coupling with 
each reservoir $E_\alpha$ is associated with 
heat $Q_\alpha$. Thus, we can write 
$d \mean{H_S}/dt = \dot W + 
\sum_\alpha \dot Q_\alpha$, where 
work and heat rates are,  
respectively, 
$\dot W = \langle \frac{\partial}{\partial t} H_S\rangle$ and
$\dot Q_\alpha =\langle\left[H_S, H_{int,\alpha}\right]  
\rangle/i$ (both work and heat are  
positive when energy flows into the network). 

We restrict to consider a periodic driving 
for which there is a stable stationary 
regime (see below).  
In this regime, the quantum state of $S$ 
is periodic and the 
average of $d\mean{H_S}/dt$ vanishes. Then, 
the asymptotic value of the 
average work in a cycle, defined as 
$\dot {\bar W} = \lim_{n\to\infty} \int_{n\:\tau}^{(n+1)\tau} 
\dot W(t) dt/\tau$ and the analogously defined  
averaged heat rate $\dot {\bar Q}_\alpha$, satisfy the identity  
$\dot {\bar W} + \sum_\alpha \dot {\bar Q}_\alpha=0$. This  
is nothing but the first law of thermodynamics
for a cyclic process. Notice that  
heat could be defined 
differently: Thus, the rate of change 
of the energy of $E_\alpha$ 
is $\dot Q'_\alpha= 
\langle\left[H_{E,\alpha}, H_{int,\alpha}\right]  \rangle/i$. 
As the interaction $H_{int}$ does not commute 
with the total Hamiltonian $H_T$, the 
energy lost by $E_\alpha$ is not equal 
to the one absorbed by $S$ (and therefore
$\dot Q_\alpha+\dot Q'_\alpha\neq 0$), 
However, as shown in SM 
\ref{ap:work_and_heat}, in the 
stationary regime, in average, 
no energy is stored in the interaction terms
and the identity $\dot {\bar Q}_\alpha+\dot {\bar Q}'_\alpha=0$
holds. 

{\it The solution:} We sketch the method
we use and leave most details for the 
Supplementary Material (SM).  
The effect of the environments on
the network depends 
on the so called spectral density $I(\omega)=
\sum_\alpha I_\alpha(\omega)$ where, for each 
reservoir, we define the matrix 
$[I_\alpha(\omega)]_{j,k}= \sum_l
 \frac{1}{m\omega}C_{\alpha,jl}
C_{\alpha,kl}\delta(\omega-\omega_{\alpha,l})$. For
example, the effect of the environment on the 
retarded Green function $g(t,t')$, 
that determines the response of the network
to an impulse and is 
a central ingredient in our analysis, shows up in 
its evolution equation which reads as 
 $\ddot g(t,t')+V_R(t)g(t,t')+\int_0^td\tau\ 
\gamma(t-\tau)\dot g(\tau,t')=\delta(t-t')
\mathds{1}$, where a dot denotes the
derivative with respect to the first temporal argument. 
The reservoirs determine the  
damping kernel $\gamma(t)$ which is  
$\gamma(t)=\int_0^\infty d\omega\ I(\omega)\cos(\omega t)/\omega$ 
and the renormalized potential $V_R(t)=V(t)-\gamma(0)$. 

We will make the 
following assumptions: a) The external driving
is periodic with period $\tau = 2\pi/\omega_d$ (i.e. 
$V_R(t) = \sum_{k=-\infty}^{+\infty} V_k
e^{ik\omega_d t}$) and 
that $\omega_d$ is such that the driving  
does not induce instabilities via parametric resonance. 
b) The coupling with the reservoirs is such that the 
Green function decays exponentially with $(t-t')$. In such conditions 
there is a stationary regime, where all  
relevant quantities are also $\tau$-periodic, 
and Floquet theory enables the 
exact solution of the problem. Details are 
shown in the Supplementary Material (SM)  \ref{ap:model},
but the crucial observation is that, for a periodic
driving, the function $p(\omega,t)=
\int_0^t g(t,t') e^{i\omega(t'-t)}dt'$ is also
$\tau$-periodic for large $t$ and therefore can be expanded
as $p(\omega,t)=
\sum_k A_k(\omega) e^{ik\omega_d t}$. The 
Fourier coefficients $A_k(\omega)$ satisfy a
set of equations obtained directly
from the one for $g(t,t')$, which read:
\begin{equation}
    \hat g\left(i(\omega +k \omega_d )\right)^{-1} A_k(\omega) + 
    \sum_{j \neq k } V_j
    A_{k-j}(\omega) = \mathds{1} \delta_{k,0}.
    \label{meq:rel_Ak}
\end{equation}
Here, 
$\hat g(s)= (M s^2 + V_0 + s \hat \gamma(s))^{-1}$, is the Laplace transform of the 
Green function of the undriven network and 
$\hat \gamma(s) = \int_0^\infty d\omega 
\; I(\omega)/\omega \;  s/(\omega^2 + s^2)$
is the transform of $\gamma(t)$ 
(see \cite{arrachea2012,arrachea2013} for a related approach). Finally, we assume: c) that the 
initial 
state has no correlations between $S$ and $E$ 
and that each reservoir is initially in a  
thermal state with temperature $T_\alpha$. 
Then, independently of 
the initial conditions, the asymptotic
state of $S$ has a Gaussian 
density matrix that 
is fully characterized by a 
$\tau$-periodic covariance matrix 
whose coefficients 
admit a Fourier series expansion. For example,
position-momentum correlation function, defined as 
$\sigma^{xp}(t) = \mean{XP^T+PX^T}(t)/2$ can be expressed as
$\sigma^{xp}(t) = \Im\left[\sum_{j,k} \sigma^{xp}_{j,k} 
\ e^{i\omega_d(j-k)t}  \right]$. Here, the Fourier components 
can be written, as 
shown in SM \ref{ap:model}, as:
\begin{equation}
    \sigma^{xp}_{j,k} = \frac{1}{2} \int_0^\infty d\omega \; (\omega+k\omega_d) \;
    A_j(\omega) \tilde \nu(\omega) A_k^\dagger(\omega) 
    \label{meq:coeff_asymp_sigma_xp},
\end{equation}
where   
$\tilde \nu(\omega) = 
\sum_\alpha I_\alpha (\omega)
\coth\left(\frac{\omega}{2 T_\alpha}\right)$ 
is the Fourier transform of 
the so called noise kernel ($\hbar=k_b=1)$. 

{\it Heat rate:}
In the asymptotic regime,  
$\dot {\bar Q}_\alpha$ is the  
temporal average of 
$\Tr \left[ P_\alpha V(t) \sigma^{xp}(t) M^{-1}\right]$, 
where $P_\alpha$ is a projector over the sites of the network
in contact with $E_\alpha$ (see 
SM \ref{ap:work_and_heat}). Using this, we 
find an important exact result: $\dot {\bar Q}_\alpha$ is the sum of three terms
\begin{equation}
    \dot{\bar Q}_\alpha =
    \dot{\bar Q}_\alpha^{\text{RP}}+
\dot{\bar Q}_\alpha^{\text{RH}}+
\dot{\bar Q}_\alpha^{\text{NRH}}.
\label{meq:local_heat}
\end{equation}
The first term, $\dot{\bar Q}_\alpha^{\text{RP}}$, 
is responsible
for the resonant pumping (RP)
of energy from (or into) $E_\alpha$. It  reads:
\begin{equation}
\begin{split}
    \dot{\bar Q}_\alpha^{\text{RP}} &= 
    \sum_{\beta \neq \alpha} \sum_k \int_{0'}^\infty d\omega \; \omega \;
    p^{(k)}_{\beta,\alpha}(\omega) \; N_\alpha(\omega)-\\
    & - \sum_{\beta \neq \alpha} \sum_k \int_{0'}^\infty
    d\omega \; (\omega+k \omega_d) \;
    p^k_{\alpha,\beta}(\omega) \; N_\beta(\omega) \; 
    \label{meq:heat_rp}
\end{split}
\end{equation}
where $N_\alpha(\omega) = (e^{\omega/T_\alpha}-1)^{-1}$ is the Planck
distribution and 
$p_{\alpha,\beta}^{(k)}(\omega)=\frac{\pi}{2} \Tr[I_\alpha(|\omega+k
\omega_d|) A_k(\omega) I_\beta(\omega) A_k^\dagger(\omega)]$ is a positive number, 
proportional to the 
probability for the network to couple the 
mode with frequency $\omega$ in 
$E_\beta$ with the one with 
frequency $|\omega+k\omega_d|$ in 
$E_\alpha$. The first term in 
Eq. (\ref{meq:heat_rp}) 
is positive and accounts for  
energy flowing out of $E_\alpha$: 
a quantum of energy $\omega$ is lost in $E_\alpha$
and excites the mode   
$|\omega+k\omega_d|$  in $E_\beta$ after 
absorbing energy $k\omega_d$ from the driving. 
The second term corresponds to the 
opposite effect: A quantum of energy $\omega$
is lost from $E_\beta$ and dumped into 
the mode $|\omega+k\omega_d|$ in $E_\alpha$
after absorbing energy $k\omega_d$ from the
driving.  

The second term in Eq. (\ref{meq:local_heat}) is
responsible for the 
resonant heating (RH) of $E_\alpha$ 
and reads: 
\begin{equation}
    \dot{\bar Q}_\alpha^{\text{RH}} = 
     - \sum_{\beta \neq \alpha} \sum_k \int_{0'}^\infty
    d\omega \; k \omega_d  \;
    p^k_{\alpha,\alpha}(\omega) \; N_\alpha(\omega) . 
    \label{meq:heat_rh}
\end{equation}
Its interpretation is analogous to the previous one
except for the fact that in 
this case energy is transferred, through the network, between modes $\omega$ and 
$|\omega+k\omega_d|$ 
in $E_\alpha$. In all cases $E_\alpha$ can gain
or loose energy $k\omega_d$, depending on the 
sign of $k$. However, when $N_\alpha(\omega)$
monotonically decreases with $\omega$ 
(as it does
for the Planck distribution) the upwards flow of
energy is more probable than the downwards
one and $E_\alpha$ heats up. 
A subtlety should be noticed: 
The lower limit of the frequency integral 
in both resonant terms (RP and RH) is not $\omega=0$
but $\omega=0'=max\{0,-k \omega_d \}$. This is 
because processes with negative $k$, which
correspond to the emission into the driving, can only 
take place if $\omega+k\omega_d>0$. 
The role of the low frequency modes (with  
$\omega<-k\omega_d$) is crucial, as we now
discuss. 

The term $\dot {\bar Q}_\alpha^{\text{NRH}}$ 
corresponds to a non resonant heating (NRH) 
effect and reads:
\begin{equation}
\begin{split}
    & \dot{\bar Q}_\alpha^{\text{NRH}} = - \sum_{k>0} \int_{0}^{k\omega_d}
     d\omega \; k \omega_d \; p_{\alpha,\alpha}^{-k}(\omega) \;
     \left(N_\alpha(\omega)\!+\!\frac{1}{2}\right) -\\
     & -\sum_{\beta \neq \alpha} \sum_{k>0} \int_{0}^{k\omega_d}
     d\omega \; (k\omega_d\!-\!\omega)\;p_{\alpha,\beta}^{-k}(\omega) \;
     \left(N_\beta(\omega)\!+\!\frac{1}{2}\right)-\\
     & -\sum_{\beta \neq \alpha} \sum_{k>0} \int_{0}^{k\omega_d}
     d\omega \; \omega \; p_{\beta,\alpha}^{-k}(\omega) \; 
     \left(N_\alpha(\omega)\!+\!\frac{1}{2}\right)
        \label{meq:heat_nrh}
\end{split}
\end{equation}
(strictly, this formula is valid if the driving is  time reversal invariant, i.e, if $V(t)=V(-t)$; 
the general expression is included in SM). 

The physical process that gives rise to 
$\dot {\bar Q}_\alpha^{\text{NRH}}$ is rather different
from the resonant ones discussed above. 
In the non resonant case a pair of excitations is 
created. One of them has energy $\omega$ while
the other one has energy  
$k\omega_d-\omega$ (these values add up
to the driving energy $k\omega_d$, notice that only
$k>0$ enters in the above expression). As opposed
to the resonant case, in NRH,
excitations are not transferred between
modes but created in pairs from the driving. 
The three terms in Eq,   
(\ref{meq:heat_nrh}) respectively correspond
to the following three cases:
i) both excitations are created
in $E_\alpha$; ii) mode $\omega$ is excited in 
$E_\alpha$ and mode $k\omega_d-\omega$ in $E_\beta$; 
iii) mode $k\omega_d-\omega$ is excited 
in $E_\alpha$ while mode $\omega$ is excited in $E_\beta$. 
In all cases $E_\alpha$ gains energy: 
in the first case the net gain is $k\omega_d$ while 
in the last two $E_\alpha$ gains, respectively, 
$\omega$ and $k\omega_d-\omega$. 
As $\dot{Q}_\alpha^{\text{NRH}}<0$, $E_\alpha$
heats up. Noticeably, the non resonant heating
term is the only one surviving when all 
reservoirs are at zero temperature (in that case
both $\dot{\bar Q}_\alpha^{\text{RP}}$ and $\dot{\bar Q}_\alpha^{\text{RH}}$ vanish).

In the absence of driving we have
$\dot{\bar Q}_\alpha^{\text{NRH}}=
\dot{\bar Q}_\alpha^{\text{RH}}=0$ and 
only $\dot{\bar Q}_\alpha^{\text{RP}}$ survives. 
In this case, only $k=0$ contributes to Eq.
(\ref{meq:heat_rp}) and the
transition probabilities are symmetric (i.e., 
 $p^{(0)}_{\alpha,\beta}(\omega)
=p^{(0)}_{\beta,\alpha}(\omega)$). 
Then, 
$\dot Q_\alpha\ge 0$ when $E_\alpha$ is the 
hottest reservoir (accordingly, 
$\dot Q_\alpha\le 0$  for the coldest one). 
This is nothing but Kelvin's version of the second
law (i.e., ``heat flows from hot to cold''
\cite{martinez2013}). 
In the driven case, Eq  
(\ref{meq:local_heat}) 
can be used to derive the most 
general form of
the second law, that reads  
$\sum_\alpha \dot {\bar Q}_\alpha/T_\alpha\le 0$. 
We omit the proof here (but
include it in SM \ref{ap:second_law} for 
completeness) and focus on the derivation 
of the third law. 

{\it Cooling and the third law:}
Since both 
$\dot{\bar Q}_\alpha^{\text{RH}}\le 0$ and $\dot{\bar Q}_\alpha^{\text{NRH}}\le 0$, the only way to 
pump energy out of $E_\alpha$
(and therefore
cool it) is to design a process such that  $\dot{Q}_\alpha^{\text{RP}}>0$. In fact, 
to pump more heat out of 
$E_\alpha$ than the 
one flowing into it, requires us to 
impose spatial (or temporal)
asymmetries to the driving or to the 
coupling with the reservoirs
\cite{levy2012,ticozzi2014,arrachea2012}. 
However, to prove the validity of the third law 
it is not necessary to go into the details of
cooling processes. For this, 
one should simply notice 
that both resonant terms vanish at
zero temperature while the non resonant heating
still survives. This alone is enough to 
establish the validity of Nernst unattainability 
principle: heating dominates 
at sufficiently low temperatures.  
Moreover, this also implies that for any cooling
process there is a minimum achievable temperature
which can be estimated as the one for 
which the cooling term in 
$\dot{\bar Q}_\alpha^{\text{RP}}$ 
becomes comparable to 
the $\dot{\bar Q}_\alpha^{\text{NRH}}$. The 
minimal temperature is not universal and its 
properties for various cooling mechanisms will be
analyzed elsewhere. Here
we focus on analyzing an important example. 
We estimate the minimal temperature that can 
achieved by the cooling strategy 
proposed in \cite{kolavr2012} 
to violate the Nernst unattainability
principle.

To gain some intuition on the behavior
of the heat rates we use the following
simplifying assumptions: a) We assume
that the driving is
weak (i.e. if $|V_k|/|V_0|\ll 1$). Then, using
perturbation theory we find that 
$A_k(\omega) = -\hat 
g(i(\omega+k\omega_d)) V_k \hat g(i\omega)$, 
for $k\neq 0$. b) We assume that 
the coupling with the
environments is also weak. Then, 
$\hat g(s)$ (the Green 
function of the undriven network) 
can be expressed as a sum over the normal 
modes of the isolated network, whose eigenfrequencies
we denote as $\Omega_a$. Using this, the 
frequency integrals in Eqs. (\ref{meq:heat_rp}) and
(\ref{meq:heat_rh}) can be performed since
the integrand is strongly peaked around 
the eigenfrequencies and their 
sidebands $\Omega_a\pm k\omega_d$ (the peaks 
in $p_{\alpha,\beta}^{k}(\omega)$ arise because
of their dependence on $A_k(\omega)$). In this
way, as shown in SM \ref{ap:weak_coupling}) the 
resonant heat rates can be expressed as a 
sum over all normal modes and sidebands. 
Finally, in the low temperature limit, this sum is
dominated by the term with the 
lowest frequency ($\Omega_0$) because the 
Planck distribution enforces a natural cutoff. 
c) To simplify the analysis
we will consider all reservoirs at the same
temperature $T_0$ (the most favorable 
condition
for cooling), and use the harmonic
driving $V(t)=V_0+2V_1
\cos(\omega_d t)$ (for which only $k=\pm 1$ appear
in the weak driving limit). In this case we obtain
\begin{equation}
\begin{split}
\dot {\bar Q}_\alpha^{\text{RP}} = & 
\frac{ e^{-\frac{\Omega_0 - \omega_d}{T_0}}
|V^{0}_1|^2 (\pi^2/8)}{\Gamma_0\Omega_0^2
(\Omega_0^2\!-\!(\Omega_0\!-\!\omega_d)^2)^2}\sum_{\beta\neq\alpha}
I_\beta^{0}(\Omega_0) I_\alpha^{0}(\Omega_0\!-\!\omega_d)\\
&\times\left\{(\Omega_0-\omega_d)-\Omega_0
\frac{I_\alpha^{0}(\Omega_0)I_\beta^{0}(\Omega_0\!-\!\omega_d)}
{I_\beta^{0}(\Omega_0)
I_\alpha^{0}(\Omega_0\!-\!\omega_d)}\right\}
   \label{meq:cooling_wwc}
 \end{split}
\end{equation}
where $\Gamma_0$ is the decay width
of the mode $\Omega_0$ (see below) and 
$M^{0}$ denotes the matrix
element of $M$ in the normal mode with frequency $\Omega_0$. 
The above equation is revealing: when the spectral densities
are identical the heat rate is  
negative and $E_\alpha$ absorbs energy. 
However, if the condition 
$I_\alpha^{0}(\Omega_0)
\ll I_\beta^{0}(\Omega_0)$ is satisfied 
$\dot{\bar Q}_\alpha^{\text{RP}}>0$ 
and the reservoir loses energy. 
This cooling condition 
simply states that the cooling process (i.e., 
extracting energy $\Omega_0 - \omega_d$ 
from $E_\alpha$ and dumping it in the 
mode $\Omega_0$ of $E_\beta$)
has a higher rate than the heating process 
(taking energy $\Omega_0-\omega_d$ 
from $E_\beta$ and dumpling it in 
mode $\Omega_0$ in $E_\alpha$). The reduction 
in the heating rate arises because 
the density of final states is small.  
As explained in SM \ref{ap:weak_coupling}, the same codition implies 
that
$\dot{\bar Q}_\alpha^{\text{RP}} \gg \dot{\bar Q}_\alpha^{\text{RH}}$.

Even if the cooling condition is satisfied, 
the heat rate rapidly decreases with the 
temperature. This is because the 
thermal factor appearing Eq. 
(\ref{meq:cooling_wwc})
decreases with temperature faster 
than any power law.
However, there is an interesting 
strategy that we could use to maximize
the heat rate. For this, as suggested in 
\cite{kolavr2012}, we can use an 
adaptive method by slowly adjusting
the driving frequency $\omega_d$ in 
such a way 
that $\Omega_0-\omega_d=T_\alpha$. In this
case, for $T_\alpha \ll \Omega_0$, the 
heat rate is 
\begin{equation}
    \dot {\bar Q}_\alpha^{\text{RP}} = \frac{\pi^2}{8e}
\frac{
 |V^{0}_1|^2 }{\Omega_0^6} T_0
I_\alpha^{0}
(T_0)\sum_{\beta\neq\alpha}
\frac{I_\beta^{0}(\Omega_0)}{\Gamma_0} 
   \label{meq:cooling_adapt}
\end{equation}
It is clear that the ratio $\sum_\beta I_\beta(\Omega_0)
/\Gamma_0$ is of zero--th order in the coupling
strength between the system and the environment. 
Therefore, the above identity shows that 
$\dot {\bar Q}_\alpha^{\text{RP}}$ is of first order in the coupling
strength. 
In contrast, it can be seen from Eq. (\ref{meq:heat_nrh})
that $\dot {\bar Q}_\alpha^{\text{NRH}}$ is of second order in the coupling
strength for $\omega_d < \Omega_0$ (since in that case the integration domain
does not include any resonance peak). 
We will use these results to estimate
the minimal temperature that can be 
achieved by this cooling protocol. For this, 
we must study cooling as a dynamical
process. 

Only a finite reservoir can be cooled. Thus, 
we assume that the 
$E_\alpha$ has a finite heat 
capacity $C_v$. As discussed in
\cite{kolavr2012}, $C_v$ depends
on the dimensionality ($d$) of the reservoir
and scales with temperature as 
$C_v \propto T_\alpha^d$.
If the rate at which energy flows
away from $E_\alpha$ is sufficiently 
small, one can think that the environment
has a time 
dependent temperature $T_\alpha(t)$
that satisfies the equation $\dot T_\alpha
=-\dot{\bar Q}_\alpha/C_v$. To solve 
this equation we need an expression for
the heat rate. We could use 
Eq. (\ref{meq:cooling_adapt}) provided 
that the rate of change of $T_\alpha$
is smaller than the driving frequency (since 
in that case we can 
instantaneously satisfy
the adaptive condition 
$\omega_d=\Omega_0-T_\alpha$). 
In this way, we obtain that $T_\alpha(t)$
satisfies 
\begin{equation}
\frac{dT_\alpha}{dt} 
= -\frac{1}{C_v} \dot Q_\alpha \; \propto \; 
    -\gamma_0 \eta \ T_\alpha^{1+\lambda_\alpha-d}
   \label{eq:ode_T_max}
\end{equation}
where we used that for low frequencies the 
spectral density $I_\alpha^0(\omega)\propto\gamma_0
\omega^{\lambda_\alpha}$ (where $\gamma_0$
is a relaxation rate scaling quadratically
with the coupling constants between the
system and the environment while 
the exponent $\lambda_\alpha$
characterizes the environment, being $\lambda_\alpha=1$
the one corresponding to a ohmic reservoir). Above, 
$\eta$ is a constant depending on $\Omega_0$
and $V_1$. The solutions of this 
equation approach
$T_\alpha=0$ in finite time when $1+\lambda_\alpha-d<1$, 
which leads us to the surprising 
conclusion that the unattainability principle could 
be violated. 
However, this argument, 
presented in \cite{kolavr2012}, 
is not correct, 
because the above equation for 
$T_\alpha$ 
stops being
valid at suffiently low temperatures. 
In that case, 
$\dot{Q}_\alpha^{\text{NRH}}$, which
was not taken into account so far, 
becomes dominant. 

The non resonant heating cannot be
neglected in spite of the fact that 
(for $\omega_d\le\Omega_0$) it is 
proportional to $\gamma_0^2$. This
scaling with $\gamma_0$ makes this
term invisible to any treatment 
based on the weak coupling
limit (such as the 
master equation used in 
\cite{kolavr2012}, which is valid to 
first order in $\gamma_0$). On the 
contrary, the weak coupling limit
captures the resonant term given in 
Eq. (\ref{meq:cooling_adapt}), which
is first order in $\gamma_0$ (in fact, 
such expression is equivalent
to the heat rate used in
\cite{kolavr2012}).

\begin{figure}[ht]
    \centering
    \includegraphics[scale=.95]{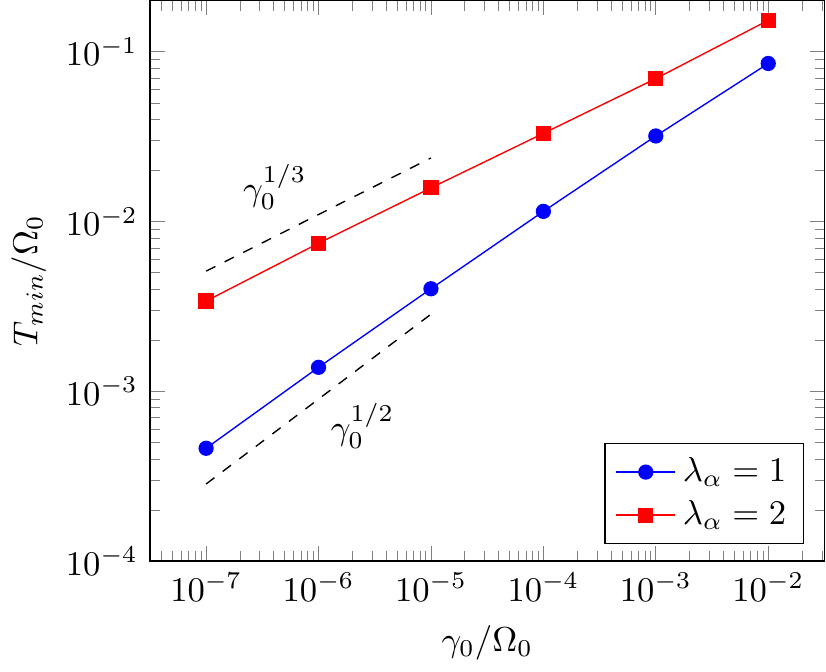}
    \caption{Minimum temperature $T_{min}$ 
    as a function of the coupling strength $\gamma_0$. $T_{min}$
    is numerically obtained as the lowest temperature for which 
    $\dot{\bar
    Q}_\alpha^{RP} > |\dot{\bar Q}_\alpha^{NRH}|$. Results are shown
    for the adaptative cooling strategy
    considered in \cite{kolavr2012}. The system consists of a single harmonic
    oscillator of frequency $\Omega_0$ in contact with two reservoirs whose
spectral densities satisfy the cooling condition  
$I_\alpha(\Omega_0)\ll I_\beta(\Omega_0)$. Heat rates are obtained
through an exact numerical evaluation (see details in SM \ref{ap:weak_coupling}).}
    \label{fig:min_temp}
\end{figure}

Our analysis, which
is non perturbative, shows that the
non resonant term given in 
Eq. (\ref{meq:heat_nrh}) will end
up stopping any cooling. 
Moreover, it enables us to 
estimate the 
minimum achievable temperature by 
estimating when resonant and 
non resonant contributions become
comparable. $\dot{\bar Q}_\alpha^{\text{NRH}}\propto
\gamma_0^2$ (and is roughly 
independent of $T_\alpha$ for sufficiently
low temperatures) and 
Eq. (\ref{meq:cooling_adapt})
shows that the cooling term scales as
$\dot{\bar Q}_\alpha^{\text{RP}}
\propto\gamma_0 
T_\alpha^{1+\lambda_\alpha}$. 
Therefore, both terms become 
comparable for temperatures scaling as
$T_\alpha\propto\gamma_0^{1/(1+\lambda_\alpha)}$. 
In Figure \ref{fig:min_temp} we see that this naive scaling 
argument is confirmed by a detailed numerical 
evaluation of both 
resonant and non resonant 
heat rates (the minimal temperature is 
estimated for various reservoirs characterized
by different values of $\lambda_\alpha$
and 
ploted as a function of $\gamma_0$). 
Thus, the dynamical 
third law (Nernst unattainability principle) has been 
restored. 

{\it Summary:} We presented a complete
description of linear quantum refrigerators
and demonstrated the validity of the third
law. Assuming linearity (i.e. a 
quadratic Hamiltonian) is not esential
for our analysis about the 
validity of such law. In fact, for
very low temperatures the discrete 
lower sector of the spectrum of the system 
is the only relevant piece of information. 
Therefore, our proof of the third law is general. 
The physical process enforcing the third law 
is the excitation of the enviromental ground
state by the driving, that ends up creating
excitation pairs in the reservoirs, as in the 
dynamical Casimir effect (whose potential relevance 
for heating, in a different setting,
was recently pointed out in \cite{benenti2015}).

This work is supported by grants from Ubacyt,
Anpcyt and Conicet (Argentina).

\bibliographystyle{unsrt}
\bibliography{article}

\onecolumngrid
\newpage
\appendix

\section{Model and analytic solution}
\label{ap:model}
We consider the following Hamiltonian,
\begin{equation}
    H = H_S + H_E + H_{int},
    \label{eq:basic_hamiltonian}
\end{equation}
where $H_S$ is the Hamiltonian of an arbitrary 
network of quantum harmonic oscillators,
and $H_E$ describes a set of uncoupled bosonic reservoirs.
$H_{int}$ is the
interaction between system and reservoirs. Therefore,
\begin{equation}
    H_S = \frac{1}{2} P^T M^{-1} P + \frac{1}{2} X^T V(t) X
    \label{eq:sys_hamiltonian},
\end{equation}
where $X$ and $P$ are vectors whose components are position and momentum
operators satisfying the usual commutation relations, $[X_i,X_j]
=[P_i,P_j]=0$ and $[X_i,P_j]=i\hbar \delta_{i,j}$. 
The matrices $M$ and $V$ are symmetric
and positive definite, and $V$ can be time-dependent.
The environmental Hamiltonian is
\begin{equation}
    H_E = \sum_\alpha H_{E,\alpha} \qquad 
    H_{E,\alpha} = \sum_{j=1}^{N_\alpha} \frac{\pi_{\alpha,j}^2}{2m} 
    + \frac{m\omega_{\alpha,j}^2}{2} q_{\alpha,j}^2
    \label{eq:env_hamiltonian}
    \nonumber
\end{equation}
That is, the environment consists of several independent reservoirs formed
in turn by a collection of quantum harmonic oscillators of mass $m$. The
operator $q_{\alpha,j}$ is the position operator of the $j$-th oscillator in the
$\alpha$-th environment, and $\pi_{\alpha,j}$ its associate momentum.
We consider a bilinear interaction between system and reservoirs through the
position coordinates
\begin{equation}
    H_{int} = \sum_\alpha \sum_{j,k} 
    C_{\alpha,jk}\; X_j\:q_{\alpha,k} 
    \label{eq:int_hamiltonian},
\end{equation}
where $C_{\alpha,jk}$ are time-independent interaction
constants. 

\subsection{Equations of motion}

We now derive, working in the Heisenberg picture, the equations of motion 
for all the operators involved in the
Hamiltonian of Eq. (\ref{eq:basic_hamiltonian}). For the system operators
the motion equations are
\begin{subequations}
\begin{align}
    \label{eq:motion_xp_sys_a}
    \dot X &= M^{-1} P  \\
    \label{eq:motion_xp_sys_b}
    \dot P &= - V(t) X - \sum_\alpha C_{\alpha} q_\alpha,
\end{align}
\end{subequations}
where $q_\alpha$ and $\pi_\alpha$ are vectors formed with the position and momentum
operators of the $\alpha$-th reservoir, respectively. Similarly, the matrix
$C_\alpha$ has as elements the interaction constants
$C_{\alpha,jk}$. If the system has $N$ degrees of 
freedom and the $\alpha$-th reservoir is formed by $N_\alpha$ oscillators, then the
matrix $C_\alpha$ has dimensions $N \times N_\alpha$.

Turning to a description in the phase space, we define the $2N$-component 
vectors $Z=(X,P)^T$ and $z_\alpha=(q_\alpha,\pi_\alpha)^T$. In terms of $Z$, the Eqs.
(\ref{eq:motion_xp_sys_a}) and (\ref{eq:motion_xp_sys_b}) can be written as:
\begin{equation}
    \dot Z + a_s(t) Z = \sum_\alpha C_\alpha z_\alpha,
    \label{eq:motion_z_sys}
\end{equation}
where the $2N\times2N$ matrix $a_s(t)$ is defined as
\begin{equation}
    a_s(t) = \begin{pmatrix}0 & -M^{-1}\\ V(t) & 0\end{pmatrix}
    \label{eq:a_s},
\end{equation}
and the $2N\times N_L$ matrix $C_L$ is
\begin{equation}
    C_\alpha = \begin{pmatrix}0 & 0\\ -C_\alpha & 0\end{pmatrix}
    \label{eq:C_l}
\end{equation}

For the operators corresponding to the $\alpha$-th reservoir the equations of
motion in phase space are:
\begin{equation}
    \dot z_\alpha + a_\alpha(t) z_\alpha = \sum_\alpha \bar C_\alpha Z
    \label{eq:motion_z_res}
\end{equation}
In this case $a_\alpha(t)$ is the $2N_\alpha\times2N_\alpha$ matrix
\begin{equation}
    a_\alpha(t) = \begin{pmatrix}0 & -\mathds{1}_{N_\alpha}/m\\ m\Omega_\alpha^2 & 0\end{pmatrix}
    \label{eq:a_l},
\end{equation}
where $\Omega_\alpha^2$ is a diagonal matrix containing the squared frequencies
of the oscillators of the $\alpha$-th reservoir. Finally, $\bar C_\alpha$ is:
\begin{equation}
    \bar C_\alpha = \begin{pmatrix}0&0\\-[C_\alpha]^T&0\end{pmatrix}
    \label{eq:bar_C_l}
\end{equation}

In summary, if we are interested in the dynamics of the system, the
following set of linear coupled differential equations 
must be solved for $Z$:
\begin{equation}
\left\{
\begin{aligned}
    \dot Z + a_s(t) Z &= \sum_\alpha C_\alpha z_\alpha, \\
    \dot z_\alpha + a_\alpha(t) z_\alpha &=  \bar C_\alpha Z
    \label{eq:motion_coupled}
\end{aligned}
\right.
\end{equation}
with the initial conditions $Z(t=0)=(X(0),P(0))^T$
and $z_\alpha(t=0)=(q_\alpha(0),\pi_\alpha(0))^T$.

\subsection{An integro-differential equation for the system}

In this section we derive an integro-differential equation describing the
dynamics of the system only. We start by considering the Green's function 
$g_\alpha(t,t')$ for the homogeneous equation of motion of the $\alpha$-th reservoir.
Such function satisfies:
\begin{equation}
    \frac{d}{dt}g_\alpha (t,t')+ a_\alpha(t) g_\alpha(t,t') =
    \mathds{1}_{N_\alpha} \delta(t-t').
    \label{eq:motion_z_res_homo}
\end{equation}
With initial conditions $g_\alpha(t'^-,t')=0$, the function $g_\alpha(t,t')$ encodes
the response of the $\alpha$-th reservoir to a delta impulse at time $t'$. For 
this simple case it just represents a rotation in phase space: $g_\alpha(t,t')
= \theta(t-t') e^{-a_\alpha(t-t')}$
where $\theta$ is the Heaviside step function and 
\begin{equation}
    e^{-a_\alpha t} = 
\begin{pmatrix}
    \cos(\Omega_\alpha t) &  \sin(\Omega_\alpha t) (m\Omega_\alpha)^{-1}\\
    - \sin(\Omega_\alpha t) m\Omega_\alpha & \cos(\Omega_\alpha t)  
\end{pmatrix}
\label{eq:green_res}
\end{equation}

The function $g_\alpha(t,t')$ is an homogeneous solution of Eq.
(\ref{eq:motion_z_res}) for $t>t'$. A particular solution is 
$z_\alpha^p(t)= \int_0^t g_\alpha(t,t') \bar C_\alpha z(t') dt'$. 
Therefore, the complete solution of Eq. (\ref{eq:motion_z_res}) is
\begin{equation}
    z_\alpha(t) = g_\alpha(t,0) z_\alpha(0) + \int_0^t g_\alpha(t,t') \bar C_\alpha z(t') dt',
    \label{eq:sol_z_res}
\end{equation}
which satisfies the required initial condition.

The solution for $z_\alpha$ of Eq. (\ref{eq:sol_z_res}) can now be inserted in Eq.
(\ref{eq:motion_z_sys}), the differential equation for the 
system coordinates. Doing this we obtain:
\begin{equation}
    \dot Z + a_s(t) Z - \int_0^t \left[ \sum_\alpha C_\alpha g_\alpha(t,t')
        \bar C_\alpha \right] Z(t') dt' =
      \sum_\alpha C_\alpha g_\alpha(t,0) z_\alpha(0)
    \label{eq:int_diff_eq}
\end{equation}
This is a non-Markovian equation of motion for the system with a source 
term depending on the operators of the environment at the initial time.
Since the equation is linear a general solution can be obtained in terms of
the Green's function of the homogeneous system, as we did before for each
reservoir. Before doing that, we define the dissipation kernel.

\subsection{Dissipation kernel}

The quantity multiplying $Z$ in the integrand of Eq. (\ref{eq:int_diff_eq})
is the dissipation kernel, to which we will refer as $\eta(t,t')=\sum_\alpha
\eta_\alpha(t,t')$, where $\eta_\alpha(t,t') =  C_\alpha g_\alpha(t,t') \bar
C_\alpha$.
It can be written in a more convenient way in
terms of the spectral densities of the reservoirs, which are defined below.
Explicitly, we have:
\begin{equation}
    \eta_\alpha(t,t') = \theta(t-t')
    \begin{pmatrix}
        0 & 0\\
        \eta_\alpha^{xx}(t-t') & 0
    \end{pmatrix}
\end{equation}
where the matrix $\eta_\alpha^{xx}$ is defined as:
\begin{equation}
    \eta_\alpha^{xx}(t)=\int_0^\infty I_\alpha(\omega) \sin(\omega t) d\omega
\end{equation}
The function $I_\alpha(\omega)$ is the spectral density associated to the $\alpha$-th reservoir. 
It is defined as follows:
\begin{equation}
 [I_\alpha(\omega)]_{j,k}= \sum_{p=1}^{N_\alpha} 
 \frac{1}{m\omega}[C_\alpha]_{jp}
 [C_\alpha]_{kp}\delta(\omega-\omega_{\alpha,p})
\end{equation}

\subsection{Solution of the equation of motion}

Using the previously defined dissipation kernel the equation of motion for 
the system is:
\begin{equation}
    \dot Z + a_s(t) Z - \int_0^t \eta(t,t')
    Z(t') dt' = \sum_\alpha C_\alpha g_\alpha(t,0) z_\alpha(0)
    \label{eq:int_diff_eq_diss}
\end{equation}
We consider the Green function $G(t,t')$ associated with the previous 
equation. It is such that:
\begin{equation}
    \frac{\partial}{\partial t} G(t,t') + a_s(t) G(t,t') - \int_0^t \eta(t,\tau)
    G(\tau,t') d\tau = \mathds{1}_{2N}\delta(t-t')
    \label{eq:int_diff_green}
\end{equation}
with initial conditions $G(t^{'-},t')=0$. 
The function $G(t,t')$ is therefore
the response of the system to an impulse at time $t'$. It fully takes into
account the non-Markovian nature of the dynamics and the dissipation induced
by the environment. In some cases the 
function $G(t,t')$ can be computed analytically. In general only
a numerical approach is possible. In any case, if $G(t,t')$ is known, 
the complete
solution to Eq. (\ref{eq:int_diff_eq_diss}) can be obtained.
In fact, it is easy to verify that the expression
\begin{equation}
    Z(t) = G(t,0) Z(0) + \int_0^t 
    G(t,t') \left[ \sum_\alpha C_\alpha g_\alpha(t',0) z_\alpha(0) \right] dt'
    \label{eq:general_sol}
\end{equation}
is a solution of Eq. (\ref{eq:int_diff_eq_diss}) and satisfies the required
initial condition.

\subsection{Renormalization and damping kernel}

It is useful to rewrite the integro-differential
equation in Eq. (\ref{eq:int_diff_green}) and express the non-Markovian integral
term as a functional of the velocity in phase space, $\frac{\partial}{\partial
t} G(t,t')$, instead of
$G(t,t')$. For that purpose
a partial integration must be performed, with the following result:
\begin{equation}
    \frac{\partial}{\partial t} G(t,t') + a_R(t) G(t,t') + \int_0^t \gamma(t,\tau)
    \frac{\partial}{\partial \tau} G(\tau,t') d\tau = \mathds{1}_{2N}\delta(t-t')
    \label{eq:int_diff_green_damp}
\end{equation}
Note that the matrix $a_s(t)$, that describes the unitary dynamics of the
system, has been renormalized to $a_R(t) = a_s(t) - \gamma(0)$. The function
$\gamma(t,t')$ is such that $\eta(t,t')= \frac{\partial}{\partial
t'} \gamma(t,t')$, and is known as the damping kernel.
For $t>t'$ it can be calculated as follows:
\begin{equation}
    \gamma_\alpha(t,t') =
    \begin{pmatrix}
        0 & 0 \\
        \gamma_\alpha^{xx}(t-t') & 0
    \end{pmatrix}
\end{equation}
where the matrix $\gamma_\alpha^{xx}$ is defined as:
\begin{equation}
    \gamma_\alpha^{xx}(t)=\int_0^\infty \frac{I_\alpha(\omega)}{\omega} \cos(\omega t) d\omega
\end{equation}

\subsection{Evolution of the covariance matrix}

The covariance matrix of the system at time $t$ is defined as 
\begin{equation}
    C(t)=\Re\left[ \langle Z(t)Z(t)^T\rangle \right] -\langle Z(t)\rangle\langle Z(t)^T\rangle
    \label{eq:def_cov}
\end{equation}
where $\langle A(t) \rangle = Tr(\rho^0 A(t))$, and $\rho^0$ is the
initial state of the system and reservoirs.We will consider initial states such that $\langle Z(0) \rangle =
\langle z_l(0) \rangle = 0$ and therefore, according to Eq. 
(\ref{eq:general_sol}),
$\langle Z(t) \rangle = 0 $ for all $t$.
 Inserting Eq. (\ref{eq:general_sol})
in Eq. (\ref{eq:def_cov}) the following expression is obtained:
\begin{equation}
C(t) = G(t,0) C(0) G(t,0)^T  
+ G(t,0) \Re \left[ \langle Z(0)\beta(t)^T\rangle \right]
+ \Re \left[ \langle \beta(t)Z(0)^T\rangle\right]G(t,0)^T
+ \Re \left[ \langle \beta(t)\beta(t)^T\rangle \right]
    \label{eq:evol_cov}
\end{equation}
where $\beta(t)$ is the integral term of Eq. (\ref{eq:general_sol}),
\begin{subequations}
\begin{align}
    \label{eq:def_beta}
    \beta(t) &= \sum_\alpha \beta_\alpha(t) z_\alpha(0)\\
    \label{eq:def_beta_l}
    \beta_\alpha(t) &= \int_0^t G(t,t')  C_\alpha g_\alpha(t',0)  dt'
\end{align}
\end{subequations}
The first term in Eq. (\ref{eq:evol_cov}) is the deterministic propagation
of the initial covariance matrix given by the phase space flow $G(t,0)$.
The second and third terms are the propagation of the initial correlations 
between system and reservoirs. The last term correspond to the noise
and diffusion induced by the environment on the system, and 
in a stable system will dominate the long term behaviour. 
If there are no system-reservoirs correlations in the initial state, 
i.e, if $\Re \left[ \langle Z(0)z_\alpha(0)^T\rangle \right] =0$
for all $\alpha$, then the second and third terms of Eq. (\ref{eq:evol_cov}) 
vanish for all $t$. In that case the evolution of
the covariance matrix is just:
\begin{equation}
C(t) = G(t,0) C(0) G(t,0)^T  + \Re \left[ \langle \beta(t)\beta(t)^T\rangle
\right]
\label{eq:evol_cov_sep}
\end{equation}

Using Eq. (\ref{eq:def_beta}) we find the following expression for the
diffusive term of the covariance matrix:
\begin{equation}
    \Re \left[ \langle \beta(t)\beta(t)^T\rangle \right]=
    \int_0^t \int_0^t 
    G(t,t_1)\left[\sum_{\alpha,\beta}C_\alpha g_\alpha(t_1,0) \Re\left[\langle
        z_\alpha(0)
    z_{\beta}(0)^T\rangle\right]
g_{\beta}(t_2,0)^T C_{\beta}^T \right] G(t,t_2)^T dt_1 dt_2
    \label{eq:diff_cov_general}
\end{equation}
Now we introduce the condition that in the initial state the reservoirs are
in thermal states and uncorrelated with each other. In that case:
\begin{equation}
    \Re \left[ \langle z_\alpha(0) z_{\beta}(0)^T\rangle \right] =
    \delta_{\alpha,\beta} 
    \frac{\hbar}{2} 
    \begin{pmatrix}
    (m\Omega_\alpha)^{-1} \coth\left(\frac{\hbar \Omega_\alpha}{2k_BT_\alpha}\right)&0\\ 
    0&(m\Omega_\alpha) \coth\left(\frac{\hbar \Omega_\alpha}{2k_BT_\alpha}\right)\\ 
    \end{pmatrix}
    \label{eq:thermal_cov},
\end{equation}
Where $T_\alpha$ is the temperature of the $\alpha$-th reservoir and $k_B$ is the
Boltzmann constant. Introducing Eq. (\ref{eq:thermal_cov}) in Eq. (\ref{eq:diff_cov_general}) the following final
expression is obtained:
\begin{equation}
    \Re \left[ \langle \beta(t)\beta(t)^T\rangle \right] = \frac{\hbar}{2}
    \int_0^t \int_0^t 
    G(t,t_1) \nu(t_1-t_2) G(t,t_2)^T dt_1 dt_2
    \label{eq:diff_cov_final}
\end{equation}
The matrix function $\nu(t) = \sum_\alpha \nu_\alpha(t)$ is the noise kernel, with
\begin{equation}
    \nu_\alpha(t) = 
    \begin{pmatrix}
        0 & 0 \\
        0 & \nu_\alpha^{xx}(t)
    \end{pmatrix}
    \label{eq:noise_kernel}
\end{equation}
where:
\begin{equation}
    \label{eq:noise_kernel_xx}
    \nu_\alpha^{xx}(t)=\int_0^\infty I_\alpha(\omega) \cos(\omega t)
    \coth\left(\frac{\hbar\omega}{2k_BT_\alpha}\right) d\omega
\end{equation}

\subsection{Asymptotic state for stable systems}

In this section we characterize the asymptotic state for driven systems 
that are exponentially stable, i.e., systems with a Green's function $G(t,t')$ 
decaying exponentially with $t-t'$. The previous condition is not always
fulfilled for driven systems, even in the presence of strong dissipation, 
since it is possible, for example, to induce a divergent dynamics by the phenomenon of 
parametric resonance. 

From Eq. (\ref{eq:evol_cov_sep}) it is clear that for stable systems the
asymptotic covariance matrix is:
\begin{equation}
    C(t) = \Re \left[ \langle \beta(t) \beta(t)^T\rangle \right]
    = \begin{bmatrix} \sigma^{xx}(t) & \sigma^{xp}(t) \\ \sigma^{px}(t) &
        \sigma^{pp}(t) \end{bmatrix}
    \label{eq:asymp_cov}
\end{equation}
since $G(t,0) \to 0$ for large $t$. 
Also, Eq. (\ref{eq:diff_cov_final}) is equivalent to the following 
expressions for each block of the asymptotic covariance matrix:
\begin{subequations}
\begin{align}
    \sigma^{xx}(t) &= \frac{\hbar}{2}
    \int_0^t \int_0^t g(t,t_1) M^{-1} \; \nu^{xx}(t_1-t_2) \; M^{-1}g(t,t_2)^T
    \; dt_1 dt_2 
    \label{eq:sigma_xx}
    \\
    \sigma^{xp}(t) &= \frac{\hbar}{2}
    \int_0^t \int_0^t g(t,t_1) M^{-1} \; \nu^{xx}(t_1-t_2) \; M^{-1}\frac{\partial}{\partial t}
    g(t,t_2)^T M \; dt_1 dt_2 
    \label{eq:sigma_xp}
    \\
    \sigma^{pp}(t) &= \frac{\hbar}{2}
    \int_0^t \int_0^t M \frac{\partial}{\partial t}g(t,t_1) M^{-1} \;
    \nu^{xx}(t_1-t_2) \; M^{-1}\frac{\partial}{\partial t} g(t,t_2)^T M \; dt_1 dt_2 
    \label{eq:sigma_pp}
\end{align}
\end{subequations}
In the previous expressions, the function $g(t,t')$ is the Green's function of
the system in the configuration space, and satisfies the following second order
integro-differential equation for $t>t'$:
\begin{equation}
    M \frac{\partial^2}{\partial t^2} g(t,t') + 
    V_R(t) g(t,t') + 
    \int_0^t \gamma^{xx}(t-\tau) \frac{\partial}{\partial \tau} g(\tau,t') d\tau = 0 
    \label{eq:greens_conf}
\end{equation}
with initial conditions $g(t=t',t') = 0$ and $\frac{\partial}{\partial t} g(t =
t',t') = \mathds{1}$. Also, $V_R(t) = V(t) - \gamma^{xx}(0)$ is the
renormalized potential. We now introduce the spectral decomposition of the noise
kernel (see Eq. (\ref{eq:noise_kernel_xx})):
\begin{equation}
    \nu^{xx}(t_1 - t_2)=\Re\left[ \int_0^\infty \tilde \nu (\omega)  e^{i\omega t_1} e^{-i\omega t_2} d\omega\right]
\end{equation}
where $\tilde \nu(\omega)$ is the Fourier transform of $\nu^{xx}(t)$:
\begin{equation}
    \tilde \nu(\omega) = \sum_\alpha I_\alpha (\omega)
    \coth\left(\frac{\hbar \omega}{2 k_b T_\alpha}\right)
    \label{eq:noise_fourier}
\end{equation}
Introducing Eq. (\ref{eq:noise_fourier}) into Eqs. (\ref{eq:sigma_xx} -
\ref{eq:sigma_pp}) we obtain:
\begin{subequations}
\begin{align}
    \sigma^{xx}(t) &= \frac{\hbar}{2}\Re\left[ 
    \int_0^\infty q(t,\omega) \; \tilde\nu(\omega) \; q(t,\omega)^\dagger
    \; d\omega \right]
    \label{eq:sigma_xx_w} \\
    \sigma^{xp}(t) M^{-1} &= \frac{\hbar}{2}\Re\left[
    \int_0^\infty q(t,\omega) \; \tilde\nu(\omega) \;
    \frac{\partial}{\partial t} q(t,\omega)^\dagger
\; d\omega\right]
    \label{eq:sigma_xp_w}\\
    M^{-1} \sigma^{pp}(t) M^{-1} &= \frac{\hbar}{2}\Re\left[ 
    \int_0^\infty \frac{\partial}{\partial t} q(t,\omega) \; \tilde\nu(\omega) \;
    \frac{\partial}{\partial t} q(t,\omega)^\dagger
    \; d\omega\right]
    \label{eq:sigma_pp_w}
\end{align}
\end{subequations}
where the function $q(t,\omega)$ is defined as:
\begin{equation}
    q(t,\omega) = \int_0^t g(t,t') e^{i\omega t'} dt'
    \label{eq:def_q}
\end{equation}
So far we have only given alternatives expressions for the asymptotic
covariance matrix valid when the system is stable.
We now analyze the asymptotic properties of the function 
$q(t,\omega)$ for the case in which the driving is periodic.

\subsection{Periodic driving}

The potential energy matrix $V(t)$ is assumed to be $\tau$-periodic:
$V(t+\tau)= V(t)$. From the integro-differential equation defining $g(t,t')$
(Eq. (\ref{eq:greens_conf})) it follows that:
\begin{equation}
    g(t+\tau,t'+\tau) = g(t,t'),
    \label{eq:g_biperiodic}
\end{equation}
since $g(t+\tau,t'+\tau)$ and $g(t,t')$ are both solution of Eq.
(\ref{eq:greens_conf}) with the same initial conditions. This observation
implies that
\begin{equation}
    q(t+\tau,\omega) = \int_0^{t+\tau} g(t+\tau,t') e^{i\omega t'} dt'
    = \int_{-\tau}^{t} g(t+\tau,t'+\tau) e^{i\omega (t'+\tau)} dt'
    = \left[ \int_{-\tau}^{t} g(t,t') e^{i\omega t'} dt' \right] \: e^{i\omega \tau}.
    \label{eq:floquet_q}
\end{equation}
Now, if $g(t,t')$ decays exponentially with $(t-t')$ the following
approximation holds for large $t$:
\begin{equation}
    \int_{-\tau}^{t} g(t,t') e^{i\omega t'} dt' \simeq
    \int_{0}^{t} g(t,t') e^{i\omega t'} dt' = q(t,\omega)
\end{equation}
Therefore, in the asymptotic limit, the function $q(t,\omega)$ satisfies:
\begin{equation}
    q(t+\tau,\omega) = q(t,\omega) e^{i\omega\tau}
\end{equation}
from which it follows that the function
\begin{equation}
    p(t,\omega) = q(t,\omega) e^{-i\omega t}
    \label{eq:def_p}
\end{equation}
is $\tau$-periodic. In summary the function $q(t,\omega)$, from which the
covariance matrix at time $t$ can be obtained, can be expressed for
sufficiently long times as $q(t,\omega)=p(t,\omega) e^{i\omega t}$, where
$p(t,\omega)$ is $\tau$-periodic. As a consequence, the asymptotic state for
long times will also be $\tau$-periodic. To see that, as an example, we rewrite
the asymptotic limit of $\sigma^{xx}(t)$ in terms of $p(t,\omega)$:
\begin{equation}
    \sigma^{xx}(t) = \frac{\hbar}{2}\Re\left[
    \int_0^\infty p(t,\omega) \; \tilde\nu(\omega) \; p(t,\omega)^\dagger
\; d\omega\right]
    \label{eq:sigma_xx_p} 
\end{equation}
similar expressions hold for $\sigma^{xp}(t)$ and $\sigma^{pp}(t)$, in which
time only enters through $p(t,\omega)$ or its derivative.

To finish this section we note that since the functions $V_R(t)$ and
$p(t,\omega)$ are $\tau$-periodic they are determined by their Fourier
coefficients $V_k$ and $A_k(\omega,\omega_d)$:
\begin{equation}
    V_R(t) = \sum_{k=-\infty}^{+\infty} V_k \: e^{i k \omega_d  t}
    \label{eq:fourier_V}
\end{equation}
\begin{equation}
    p(t,\omega) = \sum_{k=-\infty}^{+\infty} A_k(\omega,\omega_d) \: e^{i k \omega_d  t}
    \label{eq:fourier_p}
\end{equation}
where $\omega_d = 2\pi/\tau$ is the fundamental angular frequency of the
driving. In the following section we explain how to calculate the coefficients
$A_k$ given the driving coefficients $V_k$. However, if they are known,
then the asymptotic correlations can be easily obtained as:
\begin{equation}
    \sigma^{xx}(t) = \Re\left[\sum_{j,k} \sigma^{xx}_{j,k} \; e^{i\omega_d(j-k)t}
    \right]
    \label{eq:asymp_sigma_xx}
\end{equation}
where:
\begin{equation}
    \sigma^{xx}_{j,k} = \frac{\hbar}{2} \int_0^\infty A_j(\omega,\omega_d) \tilde \nu(\omega)
    A_k^\dagger(\omega,\omega_d) d\omega
    \label{eq:coeff_asymp_sigma_xx}
\end{equation}
Similar expressions can be found for the correlations $\sigma^{xp}(t)$ and
$\sigma^{pp}(t)$.

\subsection{Calculation of the function $p(t,\omega)$}

An integro-differential equation for $p(t,\omega)$ can be derived from the one
defining $g(t,t')$, Eq. (\ref{eq:greens_conf}). It reads:
\begin{equation}
    M\left[\frac{\partial^2}{\partial t^2} p(t,\omega) + 
    2(i\omega) \frac{\partial}{\partial t} p(t,\omega) +
    (i\omega)^2 p(t,\omega) \right] + V_R(t) P(t,\omega) + 
    \int_0^t \gamma^{xx}(t-\tau) \left[ \frac{\partial}{\partial t} p(\tau,\omega) + (i\omega)
    p(\tau,\omega) \right] e^{-i\omega(t-\tau)} d\tau = \mathds{1}
    \label{eq:int_diff_p}
\end{equation}
Inserting the Fourier decomposition of Eq. (\ref{eq:fourier_p}) into the
previous equation one obtains the following algebraic relation for the coefficients
$A_k(\omega,\omega_d)$:
\begin{equation}
    \hat g\left(i(\omega + k\omega_d )\right)^{-1} \: A_k(\omega,\omega_d) + \sum_{j \neq k } V_j
    A_{k-j}(\omega,\omega_d) = \mathds{1} \delta_{k,0}
    \label{eq:rel_Ak}
\end{equation}
where the matrix function $\hat g(s)$ is the Laplace transform of the Green's
function of the system without driving, which satisfies:
\begin{equation}
    \hat g(s)^{-1} = M s^2 + V_R + s \hat \gamma(s)
    \label{eq:def_laplace_green}
\end{equation}
In turn, $\hat \gamma(s)$ is the Laplace transform of the damping kernel
$\gamma^{xx}(t)$:
\begin{equation}
    \hat \gamma(s) = \int_0^\infty \frac{I(\omega)}{\omega} \frac{s}{\omega^2 + s^2}
    d\omega
    \label{eq:def_laplace_damping}
\end{equation}
The infinite set of equations given in Eq. (\ref{eq:rel_Ak}) can be solved for
any given value of $\omega$ by standard techniques. For example, a finite
linear system can be obtained by only considering coefficients
$A_k(\omega,\omega_d)$
with $|k|\leq k_{max}$, which is later solved by a regular matrix inversion.
Alternatively, a perturbative approach can be employed. Thus, if the driving is
weak (i.e., if $|V_k| \ll |V_0|$ for all $k\neq 0$), then up to second 
order in $V_k$ we have
\begin{subequations}
\begin{align}
    A_0(\omega,\omega_d) &= \hat g(i\omega) + \sum_{k\neq 0} g(i\omega) \; V_k
    \;g(i(\omega-k \omega_d ))\; V_{-k} \; g(i\omega)\\
    A_k(\omega,\omega_d) &= -\hat g(i(\omega+k \omega_d)) \; V_k\; \hat g(i\omega) \:\:\:\:
    \text{for }  k\neq0
    \label{eq:weak_Ak}
\end{align}
\end{subequations}

To finish this section, we note that the coefficients $A_k(\omega,\omega_d)$ satisfy certain
exact symmetries, which can be obtained by examining the linear system given by
Eq. (\ref{eq:rel_Ak}). Thus, if $\{A_k(\omega,\omega_d)\}$ are the solutions of 
Eq. (\ref{eq:rel_Ak}) for a given process $V(t)$, and $\{A_k^r(\omega,\omega_d)\}$
are the solutions corresponding to the time reversed process $V(-t)$, we have:
\begin{subequations}
\begin{align}
    A_k^r(\omega,\omega_d) &= A_{-k}(\omega,-\omega_d)\label{eq:symm_a}\\
    A_k^r(\omega,\omega_d) &= A_{-k}^T(\omega+k \omega_d,\omega_d)\label{eq:symm_b}\\
    A_k^*(\omega,\omega_d) &= A_{-k}(-\omega,\omega_d)\label{eq:symm_c}
\end{align}
\end{subequations}
As will be clear in the next sections, if two reservoirs are connected
to sites $\alpha$ and $\beta$ of the network, then the function
$|(A_k(\omega,\omega_d))_{\alpha, \beta}|^2$ is related to the rate at which
a quantum of energy $\hbar\omega$ is extracted from the reservoir at $\beta$
while a quantum of energy $\hbar(\omega+\omega_d)$ is dumped into the reservoir at $\alpha$ (via
absortion of $k \hbar \omega_d$ energy from the driving, for $k>0$).
Thus, the above relations express fundamental symmetries between energy
exchange processes.

\section{Definition of work and heat rates}
\label{ap:work_and_heat}

In this section we give microscopic definitions for the work performed on the
system by the driving and for the energy exchange with each thermal
reservoir, i.e, for the heat rates. We begin by analyzing the variation
in time of the system energy. From Eq. (\ref{eq:sys_hamiltonian}) we have:
\begin{equation}
    \langle H_S \rangle(t) = \frac{1}{2} \Tr\left[ M^{-1} \sigma^{pp}(t)\right]  +
    \frac{1}{2} \Tr\left[ V(t) \sigma^{xx}(t)\right]
    \label{eq:mean_energy},
\end{equation}
and therefore,
\begin{equation}
    \frac{d}{d t} \langle H_S \rangle(t) = 
    \frac{1}{2} \Tr\left[ M^{-1} \frac{d}{d t} \sigma^{pp}(t)\right]  +
    \frac{1}{2} \Tr\left[ V(t) \frac{d}{d t}\sigma^{xx}(t)\right] +
    \frac{1}{2} \Tr\left[ \frac{d}{d t}V(t) \sigma^{xx}(t)\right] 
    \label{eq:rate_mean_energy},
\end{equation}
The last term in the previous equation is the rate at which energy is injected
into or absorbed from the system by the driving. The remaining terms represent
the variation of the system energy due to the interaction with the thermal
reservoirs. In order to see that it is useful to rewrite $\frac{d}{d t} \langle
H_S\rangle (t)$ as:
\begin{equation}
    \frac{d}{d t} \left\langle H_S \right\rangle (t) =
    \frac{1}{i\hbar} \left\langle\left[H_S, H\right]  \right\rangle+
    \left\langle \frac{\partial}{\partial t}H_S
    \right\rangle 
    =\frac{1}{i\hbar} \left\langle\left[H_S, H_{int}\right]  \right\rangle+
    \frac{1}{2} \Tr\left[ \frac{d}{d t}V(t) \sigma^{xx}(t)\right] 
    \label{eq:rate_heisenberg}
\end{equation}
where $H$ is the total Hamiltonian defined in Eq. (\ref{eq:basic_hamiltonian}).
Comparing Eqs. (\ref{eq:rate_mean_energy}) and (\ref{eq:rate_heisenberg}) we
see that the first two terms in Eq. (\ref{eq:rate_mean_energy}) can be
interpreted as the energy exchange with the reservoirs. Thus, we arrive at the
following definitions for the work rate $\dot W$ and total heat rate $\dot
Q(t)$: 
\begin{equation}
    \dot W(t) = \frac{1}{2} \Tr\left[ \frac{d}{d t}V(t) \sigma^{xx}(t)\right] 
    \label{eq:def_work_rate}
\end{equation}
and,
\begin{equation}
    \dot Q(t) = \frac{1}{i\hbar} \left\langle\left[H_S, H_{int}\right]  \right\rangle
    =\frac{1}{2} \Tr\left[ M^{-1} \frac{d}{d t} \sigma^{pp}(t)\right]  +
    \frac{1}{2} \Tr\left[ V(t) \frac{d}{d t}\sigma^{xx}(t)\right] 
    \label{eq:def_heat_rate}
\end{equation}

\subsection{Local heat rates}

Equation (\ref{eq:def_heat_rate}) defines the total energy interchange
between the system and all the thermal reservoirs. However, 
the local heat rate corresponding to a particular 
reservoir is also of interest. 
A working definition for such local heat rates can be obtained by
expanding Eq. (\ref{eq:def_heat_rate}) using Eq. (\ref{eq:int_hamiltonian}):
\begin{equation}
    \dot Q = \frac{1}{i\hbar} \left\langle\left[H_S, H_{int}\right]  \right\rangle
    = \sum_{\alpha} \frac{1}{i\hbar} \left\langle\left[H_S,
    H_{int,\alpha}\right]  \right\rangle
    \label{eq:total_heat_sum}
\end{equation}
where $H_{int,\alpha} = \sum_{j,k} C_{\alpha,jk}\; X_j\:q_{\alpha,k} = X^T
C_\alpha q_\alpha$ is the Hamiltonian term describing the interaction between
the system and the $\alpha$-th reservoir. We define
\begin{equation}
    \dot Q_\alpha = \frac{1}{i\hbar} \left\langle\left[H_S, H_{int,\alpha}\right]  \right\rangle
    \label{eq:local_heat}
\end{equation}
as the heat rate corresponding to the $\alpha$-th reservoir. In this way we
obtain a set $\{\dot Q_\alpha\}$ of local heat rates such that the total
heat rate is $\dot Q = \sum_\alpha Q_\alpha$. A direct
calculation shows that
\begin{equation}
    \dot Q_\alpha = -\langle P^T M^{-1} C_\alpha q_\alpha \rangle
    \label{eq:local_heat_2}
\end{equation}
We can use the motion equation (\ref{eq:motion_xp_sys_b}) in order to eliminate
the reservoir coordinates from Eq. (\ref{eq:local_heat_2}). Thus, if
$P_\alpha$ is a projector over the sites of the network in contact with the
$\alpha$-th reservoir, then $P_\alpha \dot P = -P_\alpha V(t)X- C_\alpha
q_\alpha$ (this identity is valid in the case in which different reservoirs are
coupled to different sites of the networks, i.e., we assume that $P_\alpha
P_\beta = \delta_{\alpha,\beta} P_\alpha$ and $P_\alpha C_\beta =
\delta_{\alpha,\beta} C_\alpha$). Therefore:
\begin{equation}
    \dot Q_\alpha = \frac{1}{2} \Tr \left[ P_\alpha \frac{d}{dt} \sigma^{pp}(t)M^{-1} 
    \right] + \Tr \left[ P_\alpha V(t) \sigma^{xp}(t) M^{-1}\right]   
    \label{eq:local_heat_3}
\end{equation}

The previous definition for the local heat rates is not the only possible. 
Another natural definition for the heat rates is given by the rate of change 
of the energy of each reservoir:
\begin{equation}
    \dot Q'_\alpha=\frac{1}{i\hbar} \left\langle\left[H_{E,\alpha},
    H_{int,\alpha}\right]\right\rangle
    \label{eq:alt_local_heat}
\end{equation}
If the interactions terms were energy conserving, i.e, if it were $[H_S + H_{E,\alpha} 
,H_{int,\alpha}]=0$, then we would have $\dot Q'_\alpha + \dot Q_\alpha = 0$
and the two definitions of heat rates would be equivalent. 
Although in our model the energy
conserving condition is not fulfilled and in general $\dot Q'_\alpha + \dot
Q_\alpha \neq 0$, it is easy to see that:
\begin{equation}
    \dot Q'_\alpha + \dot Q_\alpha = \frac{d}{dt}\mean{X^tP_\alpha(\dot
    P+V(t)X)}
    \label{eq:diff_heat_defs}
\end{equation}
Since the asymptotic state is $\tau$-periodic, the right hand side of the last
equation is the derivative of a $\tau$-periodic function. This observation
implies that the average heat rates per cycle obtained with the two possible
definitions are equivalent, as if the interaction terms were energy conserving.
This is explained in the next section.

\subsection{Work and heat in the asymptotic state}

We have seen that if the system is periodically driven and stable the asymptotic 
state is also periodic, with the same period as the driving. It follows that
the function $\langle H_S \rangle(t)$ and its 
derivative $\frac{d}{dt} \langle H_S \rangle (t)$ are also periodic. 
Thus, averaging Eq. (\ref{eq:rate_mean_energy}) in one
period (for long times) we obtain:
\begin{equation}
    0 = \dot {\bar Q} + \dot {\bar W}
    \label{eq:first_law}
\end{equation}
where $\dot {\bar W}$ and $\dot {\bar Q}$ are the average work and total heat rates per cycle:
\begin{equation}
    \dot {\bar W} = \frac{1}{\tau} \lim_{n \to \infty } \int_{n\: \tau}^{(n+1)\tau} \dot W(t') \; dt' 
\qquad \qquad
    \dot {\bar Q} = \frac{1}{\tau}\lim_{n \to \infty } \int_{n \tau}^{(n+1)\tau} \dot Q(t') \; dt'
\end{equation}
Equation (\ref{eq:first_law}) is nothing more than the expression of the first
law of thermodynamics for cyclic processes. In the last equation we considered
the fact that the first term of the right hand side of Eq.
(\ref{eq:def_heat_rate}) is the derivate of a periodic function and therefore
does not contribute to the integral over a period. In the same way we can
define the local heat rates per cycle:
\begin{equation}
    \dot {\bar Q}_\alpha = \frac{1}{\tau} \lim_{k \to \infty } \int_{k\tau}^{(k+1)\tau} \dot Q_\alpha(t') dt'
    = \frac{1}{\tau} \lim_{k \to \infty } \int_{k\tau}^{(k+1)\tau}
    \Tr \left[ P_\alpha V(t') \sigma^{xp}(t') M^{-1}\right] dt'
    \label{eq:def_local_heat_cycle}
\end{equation}

This set of heat rates trivially satisfy:
\begin{equation}
    \dot {\bar Q} = \sum_\alpha \dot {\bar Q}_\alpha
    \label{eq:sum_total_heat}
\end{equation}

From Eq. (\ref{eq:diff_heat_defs}) and the fact that the right hand side is a
derivative of $\tau$-periodic function for long times it follows that:
\begin{equation}
    \dot {\bar Q'}_\alpha + \dot {\bar Q}_\alpha = 0
\end{equation}
Therefore, in a complete cycle the variation of energy of a given reservoir is
equal (in absolute value) to the variation of the energy of the system due to the
interaction with that reservoir. No energy is stored in the interaction terms.
We stress that this is true only for the averaged heat rates.

\subsection{Heat transfer matrix}

From Eqs. (\ref{eq:sigma_xp_w}), (\ref{eq:def_q}) and (\ref{eq:fourier_p}) it
is straightforward to derive the following expression for the correlation
between position and momentum in terms of the Fourier coefficients
$\left\{ A_k(\omega,\omega_d) \right\} $:
\begin{equation}
    \sigma^{xp}(t) = \Im\left[\sum_{j,k} \sigma^{xp}_{j,k} \; e^{i\omega_d(j-k)t}
    \right]
    \label{eq:asymp_sigma_xp}
\end{equation}
where:
\begin{equation}
    \sigma^{xp}_{j,k} = \frac{\hbar}{2} \int_0^\infty (\omega + k \omega_d ) A_j(\omega,\omega_d) \tilde \nu(\omega)
    A_k^\dagger(\omega,\omega_d) \;  \; d\omega
    \label{eq:coeff_asymp_sigma_xp}
\end{equation}
Introducing Eq. (\ref{eq:asymp_sigma_xp}) into the expression for the local
heat rates of Eq. (\ref{eq:def_local_heat_cycle}), and performing the time
integral, the following result is obtained:
\begin{equation}
    \dot {\bar Q}_\alpha = \frac{\hbar}{2} \int_0^\infty \Im \left\{ \sum_{j,k} \Tr \left[ P_\alpha
    V_{k-j} A_j(\omega,\omega_d) \tilde \nu (\omega)
A_k^\dagger(\omega,\omega_d) \right] 
(\omega + k \omega_d ) \right\} d\omega
\end{equation}
Now, expanding the Fourier transform of the noise kernel as in Eq.
(\ref{eq:noise_fourier}), the local heat ${\bar Q}_\alpha$ can be written as:
\begin{equation}
    \dot {\bar Q}_\alpha = \sum_\beta \int_0^\infty Q_{\alpha, \beta}(\omega)
    \coth\left(\frac{\hbar\omega}{2k_bT_\beta}\right) d\omega
    \label{eq:local_heat_trans_mat}
\end{equation}
where the functions $Q_{\alpha,\beta}(\omega)$ are defined as
\begin{equation}
    Q_{\alpha,\beta}(\omega) = \frac{\hbar}{2} \Im \left\{ \sum_{j,k} (\omega +
    k \omega_d )\Tr \left[ P_\alpha
    V_{k-j} A_j(\omega,\omega_d) I_\beta(\omega) A_k^\dagger(\omega,\omega_d) \right] 
 \right\} 
    \label{eq:def_trans_mat}
\end{equation}
If the number of reservoirs is $L$, there are $L^2$ functions
$Q_{\alpha,\beta}(\omega)$, which are considered to be the elements of matrix
called the heat transfer matrix. They specify how the heat per cycle
corresponding to the $\alpha$-th reservoir is affected by the temperature of
the $\beta$-th reservoir. The previous expression for $Q_{\alpha,\beta}(\omega)$ can be simplified. In order to do that we
note that one of the
sums appearing in Eq. (\ref{eq:def_trans_mat}) can be performed with the aid
of the algebraic equation that the coefficients $A_k(\omega,\omega_d)$ satisfy. Indeed,
from Eq. (\ref{eq:rel_Ak}) it follows that
\begin{equation}
    \sum_{j} V_{k-j} A_{j}(\omega,\omega_d) = 
    \mathds{1} \delta_{k,0} - \left[\hat g\left(i(\omega + k \omega_d )\right)^{-1}
-V_0\right]\: A_k(\omega,\omega_d) 
    \label{eq:sum_VA}
\end{equation}
Another important relation is:
\begin{equation}
    \Im\left\{\hat g\left(i\omega\right)^{-1}-V_0\right\}=
    \omega  \Re\left\{\hat \gamma(i\omega)\right\} = 
    \frac{\pi}{2} I(\omega)
    \label{eq:fluc_diss}
\end{equation}
where in the last equality the fluctuation-dissipation theorem, 
$\Re\left\{\hat\gamma(i\omega)\right\} = \frac{\pi}{2}
\frac{I(|\omega|)}{|\omega|}$, was employed.
The last equation is valid if the spectral density $I(\omega)$,
originally defined only for positive frequencies, is extended in a odd way for
negative frequencies, i.e, such that $I(-\omega)=-I(\omega)$.
Taking into account Eqs. (\ref{eq:sum_VA}) and (\ref{eq:fluc_diss}) it is
possible to arrive at the following simplified expression for the 
non-diagonal terms of the heat transfer matrix:
\begin{equation}
    Q_{\alpha,\beta}(\omega) = \frac{-\pi \hbar }{4} \sum_{k}
    (\omega + k \omega_d ) \Tr \left[ 
    I_\alpha(\omega+ k \omega_d ) A_k(\omega,\omega_d) I_\beta(\omega)
A_k^\dagger(\omega,\omega_d) \right] 
 \;\;\;\; (\alpha \neq \beta)
    \label{eq:non_diag_trans_mat}
\end{equation}
Also, the sum over the first index can be
expressed as:
\begin{equation}
    \tilde Q_\beta(\omega) = \sum_\alpha Q_{\alpha,\beta}(\omega) = \frac{-\pi \hbar }{4} \sum_{k} 
    ( k \omega_d ) \Tr \left[ 
    I(\omega+k \omega_d ) A_k(\omega,\omega_d) I_\beta(\omega)
A_k^\dagger(\omega,\omega_d) \right] 
    \label{eq:sum_first_index}
\end{equation}
The last two equations completely determine all the elements of the heat
transfer matrix. These expressions can be compared to the ones obtained in
\cite{martinez2013} for the case without driving.

\subsection{Heat rates in terms of elementary processes}

Eq. (\ref{eq:local_heat_trans_mat}) is a simple and compact expression for the
heat rates. However, as we will see, it condenses in a single formula terms with very
different physical origins, and therefore it is not possible to assign a clear
physical interpretation to each coefficient $Q_{\alpha,\beta}(\omega)$ of the heat transfer matrix. 
In this section we analyze Eqs.
(\ref{eq:local_heat_trans_mat}), (\ref{eq:non_diag_trans_mat}) and
(\ref{eq:sum_first_index}) and identify different mechanisms of heat generation
and energy transport between reservoirs. We begin by using Eq. (\ref{eq:sum_first_index}) to rewrite Eq.
(\ref{eq:local_heat_trans_mat}) as:
\begin{equation}
    \dot {\bar Q}_\alpha = \int_0^\infty d\omega \; \tilde Q_{\alpha}(\omega)
    \coth\left(\frac{\hbar\omega}{2k_bT_\beta}\right)  + 
    \sum_{\beta\neq\alpha} \int_0^\infty d\omega\; \left\{ Q_{\alpha, \beta}(\omega)
    \coth\left(\frac{\hbar\omega}{2k_bT_\beta}\right)  - 
    Q_{\beta, \alpha}(\omega) \coth\left(\frac{\hbar\omega}{2k_bT_\alpha}\right)
\right\}
\end{equation}
Expanding $\tilde Q_{\alpha}(\omega)$ and $Q_{\alpha, \beta}(\omega)$ using Eqs. (\ref{eq:non_diag_trans_mat}) and (\ref{eq:sum_first_index}) 
it is clear that some terms in the previous expression cancel out. Taking that
into account, and using the identity $\coth\left(\frac{\hbar\omega}{2k_bT_\beta}\right) =
2(N_\alpha(\omega) + 1/2)$, where
$N_\alpha(\omega)=(e^{\hbar\omega/(kT)}-1)^{-1}$ is the Planck distribution, we
write:
\begin{equation}
\begin{split}
    \dot {\bar Q}_\alpha =&-
    \sum_k \int_0^\infty d\omega \; \;\hbar k\omega_d\;
    \frac{\pi}{2}\Tr[I_\alpha(\omega+k\omega_d)A_k(\omega,\omega_d)I_\alpha(\omega)A_k^\dagger(\omega,\omega_d)]
    \; (N_\alpha(\omega)+1/2) \\ 
    &- 
    \sum_{\beta\neq\alpha} \sum_k \int_0^\infty d\omega \; \; \hbar(\omega + k\omega_d)\;
    \frac{\pi}{2}\Tr[I_\alpha(\omega+k\omega_d)A_k(\omega,\omega_d)I_\beta(\omega)A_k^\dagger(\omega,\omega_d)]
    \; (N_\beta(\omega)+1/2) \\
    &+ 
    \sum_{\beta\neq\alpha} \sum_k \int_0^\infty d\omega \; \; \hbar\omega\;
    \frac{\pi}{2}\Tr[I_\beta(\omega+k\omega_d)A_k(\omega,\omega_d)I_\alpha(\omega)A_k^\dagger(\omega,\omega_d)]
    \; (N_\alpha(\omega)+1/2)
    \label{eq:heat_aux}
\end{split}
\end{equation}

We now analyze the heat rate $\dot {\bar Q}_\alpha|_{T=0}$ when all the reservoirs are at zero
temperature, which corresponds to removing the Planck's distributions from the
previous equation. In order to simplify the discussion, we also consider the
case in which the driving protocol is invariant under time reversal (i.e, such
that $V(t)=V(-t)$). In that case the relation of Eq. (\ref{eq:symm_b}) reads
$A_k(\omega,\omega_d) = A_{-k}^T(\omega+k\omega_d,\omega_d)$. Using this
symmetry and Eq. (\ref{eq:symm_c}), via an straightforward change of variable in the integrals, 
it is possible to relate the terms with negative $k$ to the ones with 
positive $k$ in the previous equation (when all the functions
$N_\alpha(\omega)$ are equal to zero). Doing that, we obtain:
\begin{equation}
\begin{split}
    \dot {\bar Q}_\alpha|_{T=0} =&-
    \sum_{k>0} \int_0^{k\omega_d} d\omega \; \;\hbar k\omega_d\;
    p_{\alpha,\alpha}^{-k}(\omega) \; (1/2) \\ 
    &- 
    \sum_{\beta\neq\alpha} \sum_{k>0} \int_0^{k\omega_d} d\omega \; \; \hbar\omega\;
    p_{\beta,\alpha}^{-k}(\omega) \; (1/2) \\ 
    &- 
    \sum_{\beta\neq\alpha} \sum_{k>0} \int_0^{k\omega_d} d\omega \; \; \hbar(k\omega_d- \omega)\;
    p_{\alpha,\beta}^{-k}(\omega) \; (1/2)
    \label{eq:heat_aux_1}
\end{split}
\end{equation}
where we have defined the functions $p_{\alpha,\beta}^k(\omega) = \frac{\pi}{2}
\Tr[I_\alpha(|\omega + k
\omega_d|)A_k(\omega,\omega_d)I_\beta(\omega)A_k^\dagger(\omega,\omega_d)]$.
This last expression can already be interpreted in terms of pairs creation as
explained in the main text.

Note that for $k<0$ and $\omega<|k|\omega_d$ one spectral density in each terms
of Eq. (\ref{eq:heat_aux}) is evaluated at a negative frequency (recall that
extending the spectral densities to negative frequencies was necessary in Eq.
(\ref{eq:fluc_diss}) in order to obtain compact expressions). Thus, we split the
integrals for the $k<0$ case in the domains $(0,|k|\omega_d)$ and
$(|k|\omega_d,+\infty)$. The resulting expression having only the integrals in $(0,|k|\omega_d)$
is the one given in Eq. (\ref{meq:heat_nrh}) for $\dot Q_\alpha^{NRH}$ in the
main text. The remaining terms (without the $1/2$ added to the Plank's
distributions, since the $T=0$ contribution is already accounted 
by $\dot Q_\alpha^{NRH}$) are $\dot Q_\alpha^{RP} + \dot Q_\alpha^{RH}$
given in Eqs. (\ref{meq:heat_rp}) and (\ref{meq:heat_rh}) in the main text.

\section{Validity of the second law of thermodynamics}
\label{ap:second_law}

\subsection{The Planck proposition}
To investigate the validity of the second law in this context we first consider
the Planck proposition: \emph{``It is impossible to construct an engine which will work in a complete cycle, and produce no effect except the raising of a weight and cooling of a heat
reservoir''}. Translated to our setting, the previous proposition means that if the
temperatures of all the reservoirs are the same, i.e, if $T_\alpha=T_0$ for all
$\alpha$, then the work performed on the system must be positive, that is,
$\dot {\bar W} \geq 0$. In other words, it must be impossible to extract work from a single
thermal reservoir. Thus, considering Eqs. (\ref{eq:first_law}),
(\ref{eq:sum_total_heat}) and (\ref{eq:sum_first_index}), we should be able to
show that:
\begin{equation}
    \dot {\bar W} = \frac{\pi \hbar}{4} \int_0^\infty
    \sum_{k=-\infty}^{+\infty} (k \omega_d ) \Tr \left[ 
    I(\omega+k \omega_d ) A_k(\omega,\omega_d) I(\omega)
A_k^\dagger(\omega,\omega_d) \right] 
    \coth\left(\frac{\hbar \omega}{2k_b T_0}\right) d\omega \; \geq\;0
\end{equation}
The main difficulty in assessing the previous inequality is that the integrand
in the left hand side has no definite sign. It is thus
convenient to write:
\begin{equation}
    \dot {\bar W} = \frac{\pi \hbar}{4} \sum_{k>0} (k \omega_d ) (I_k(\omega_d)
    - I_{-k}(\omega_d))
\end{equation}
where the functions $I^{\pm}_k(\omega_d)$ are defined as:
\begin{equation}
    I_k(\omega_d) = \int_0^\infty \Tr \left[ 
    I(\omega + k \omega_d ) A_{k}(\omega,\omega_d) I(\omega) 
    A_{k}^\dagger(\omega,\omega_d) \right] 
    \coth\left(\frac{\hbar \omega}{2k_b T_0}\right) d\omega 
\end{equation}
Now, if $I_{k}^r(\omega_d)$ is the function corresponding to the time
reversed process $V(-t)$ it is possible to show that $I_k^r(\omega_d) \geq
I_{-k}(\omega_d)$, which we do as follows:

\begin{equation}
    I_{-k} = \int_0^\infty \dots d\omega = \int_0^{k \omega_d } \dots d\omega + 
    \int_{k \omega_d }^\infty  \dots d\omega  
\end{equation}
The first integral in the right hand side is always negative, since
$I(\omega-k\omega_d )$ is negative in the interval $(0,k\omega_d)$. The second
integral is:
\begin{equation}
\begin{split}
    &\int_{k \omega_d }^\infty \Tr \left[ 
    I(\omega-k \omega_d ) A_{-k}(\omega,\omega_d) I(\omega)
A_{-k}^\dagger(\omega,\omega_d) \right] 
    \coth\left(\frac{\hbar \omega}{2k_b T_0}\right) d\omega  = \\
    &\int_{0}^\infty \Tr \left[ 
    I(\omega) A_{-k}(\omega+k \omega_d ,\omega_d) I(\omega+k \omega_d )
A_{-k}^\dagger(\omega+k \omega_d ,\omega_d) \right] 
    \coth\left(\frac{\hbar (\omega+k \omega_d )}{2k_b T_0}\right) d\omega  \leq  \\
    &\int_{0}^\infty \Tr \left[ 
        I(\omega+k \omega_d ) {A_{k}^r}^T(\omega,\omega_d) I(\omega)
    {A_{k}^r}^*(\omega,\omega_d) \right] 
    \coth\left(\frac{\hbar \omega}{2k_b T_0}\right) d\omega  = I_k^r  \\
\end{split}
\end{equation}
The first step was a simple change of variable and in the second step we used
the identity of Eq. (\ref{eq:symm_b}) and the fact that
$\coth(\omega)$ is a decreasing function of $\omega$. Therefore, we proved that
$I_{-k} \leq I_k^r$, as required. Also, $I_{-k}^r \leq I_{k}$. It follows that
$\dot {\bar W} + \dot {\bar W}^r \geq 0$. Using Eq. (\ref{eq:symm_a}) it is easy to
see that as a function of $\omega_d$ we have $\dot {\bar W}^r(\omega_d) = \dot
{\bar W}(-\omega_d)$. In summary, we have shown that
\begin{equation}
    \dot {\bar W}(\omega_d) + \dot {\bar W}(-\omega_d) \geq 0
    \label{eq:pos_sum}
\end{equation}
We now note the following three facts: (i) the function $\dot {\bar W}(\omega_d)$ is
continuous, (ii) the only root of the function $\dot {\bar W}(\omega_d)$ is $\omega_d
= 0$ and (iii) $\dot {\bar W}(\omega_d)  = \dot {\bar W}(-\omega_d)$ for $\omega_d \to \pm
\infty$ (this last fact is physically obvious but can be verified using the
approximate solutions of the coefficients $A_k(\omega,\omega_d)$ for large
$\omega_d$). It follows that $\dot {\bar W}(\omega_d) \geq 0$ for all $\omega_d$.

\subsection{Irreversible entropy generation}
A general derivation of the second law in our setting can be obtained without
using the specific expression for the heat rates. Thus, we would like to
show that in a complete cycle of the process the production of entropy is
always positive, i.e,
\begin{equation}
    \sum_\alpha \frac{-\dot {\bar Q}_\alpha}{T_\alpha} \geq 0
    \label{eq:gen_sec_law}
\end{equation}
This statement of the second law is equivalent to the Planck principle only in
the adiabatic limit (only in that limit the heat transfer matrix in symmetric), 
as can be verified by direct calculation in the high temperature regime.
Although it must be possible in principle to prove Eq. (\ref{eq:gen_sec_law})
from the general expression for the heat transfer matrix, that is not the more economic or 
elegant approach. To prove Eq. (\ref{eq:gen_sec_law}) we begin by considering
the variations of the Von Neumann entropy of each reservoir and the system.
We take $S(t)$ and $S_{\alpha}(t)$ as the entropies of the system and the 
$\alpha$-th reservoir at time $t$. Also, $\Delta S(t) = S(t)-S(0)$. Using that
the initial global state is a product state, that the global dynamics is unitary
and the subadditivity of the Von Neumann entropy it is easy to see that:
\begin{equation}
    \Delta S(t) + \sum_\alpha \Delta S_\alpha(t) \geq 0
    \label{eq:subadd}
\end{equation}
We now divide the previous equation by the total time and evaluate in
$t=k\tau$. Since the asymptotic state of the system is $\tau$-periodic, its
entropy is also a $\tau$-periodic, and since it is continuous, it is also
bounded. Therefore $\lim_{k\to\infty} \Delta S(k\tau) / (k\tau) = 0$. On the other
hand $\Delta S_\alpha(k \tau) /(k\tau)$ converges to $\Delta
S^c_\alpha/\tau$ for $k\to\infty$,
where $\Delta S^c_\alpha$ is the change in entropy of the $\alpha$-th
reservoir per cycle in the asymptotic state. In this way, we obtain the following
inequality for the variations per cycle of the entropy of the reservoirs in the
asymptotic state:
\begin{equation}
    \sum_\alpha \Delta S_\alpha^c \geq 0
    \label{eq:asymp_subadd}
\end{equation}
We now take advantage of the fact that the initial state of the $\alpha$-th reservoir is
thermal at temperature $T_\alpha$, and
therefore it is the only minimum of the free energy function $F_\alpha(\rho) =
\Tr(\rho H_{E,\alpha}) - k_b T_\alpha S(\rho)$. As a consequence:
\begin{equation}
    0 \leq \Delta F_\alpha(t) = \Delta E_\alpha(t) - k_b T_\alpha \Delta
    S_\alpha(t)
    \label{eq:min_free}
\end{equation}
As before, dividing the previous equation by $t$, evaluating in $t=k\tau$, and
taking the limit $k \to \infty$, we obtain:
\begin{equation}
    \frac{\Delta E_\alpha^c}{k_b T_\alpha} \geq \Delta S^c_\alpha
    \label{eq:asymp_min_free}
\end{equation}
where $\Delta E_\alpha^c$ is the variation per cycle of the energy of the
$\alpha$-th reservoir in the asymptotic state. It is not immediately obvious how this variation is
related to the previously defined heat rates (since in our model the
interaction terms are not energy conserving, i.e, $[H_S + H_{E,\alpha},
H_{int,\alpha}] \neq  0 $). However, in Appendix \ref{ap:work_and_heat} we show
that in the asymptotic state $\Delta E_\alpha^c = - \tau \dot {\bar Q}_\alpha$, as
expected. Inserting this last identity in Eq. (\ref{eq:asymp_min_free}), summing over all
reservoirs, and using Eq. (\ref{eq:asymp_subadd}) we obtain Eq.
(\ref{eq:gen_sec_law}).

\section{Weak coupling approximation, minimum cooling temperature, and numerical evaluation of the heat
rates}
\label{ap:weak_coupling}

\begin{figure}[ht]
    \centering
    \includegraphics{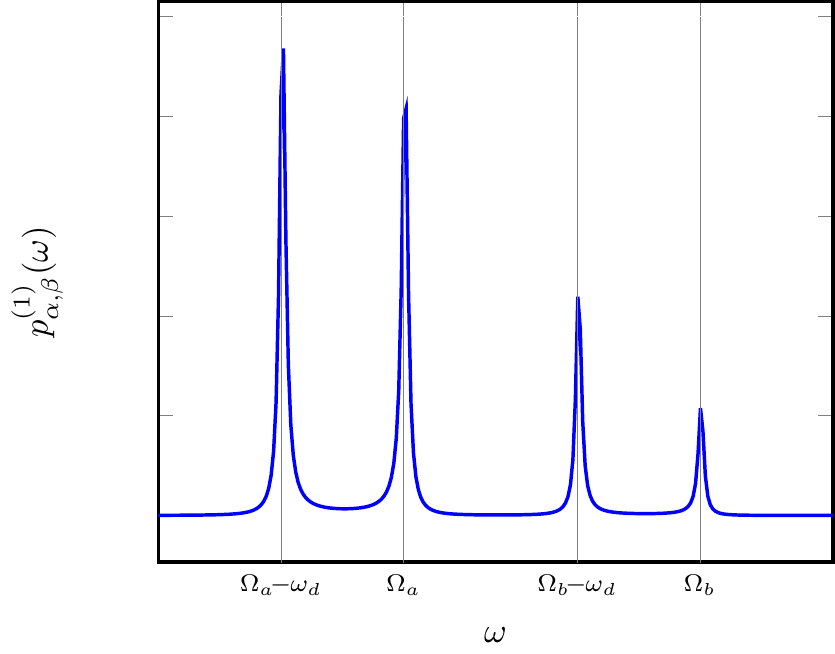}
    \caption{Typical shape of the function $p_{\alpha,\beta}^{(1)}(\omega)$ in the weak coupling limit}
    \label{fig:example_wc}
\end{figure}

In the weak coupling regime the frequency integrals in Eq. (\ref{eq:heat_aux})  
the heat transfer matrix can be approximated by sums over the 
normal modes of the closed system. To see that we need an analytic 
expression for the coefficients $A_k(\omega,\omega_d)$. Thus, we employ 
the weak driving approximation of Eq. (\ref{eq:weak_Ak}), which gives the
solution for those coefficients in terms of the Laplace's transform
of the Green's function $\hat g(i\omega)$. As explained in \cite{freitas2014},
in the weak coupling limit $\hat g(i\omega)$ can be approximated as
\begin{equation}
    \hat g(i\omega) = \sum_{a} \frac{q_a q_a^T}{\Omega_a ^2 - (w-i\Gamma_a)^2} 
    \label{eq:weak_green_nm}
\end{equation}
where $\{\Omega_a\}$ and $\{q_a\}$ are the normal 
frequencies and modes of the
closed system and $\Gamma_a$ is 
the dissipation rate of each normal mode. We assume for simplicity that 
the system is not degenerated. 
As an example, we write the expression
for $p_{\alpha,\beta}^{(k)}(\omega)$ using the
previous approximations (for $k\neq0$):
\begin{equation}
    p_{\alpha,\beta}^{(k)}(\omega) = 
    \frac{\pi}{2} 
    \sum_{a,b,c,d} 
    \frac{(q_a^T V_k q_b) (q_d^T V_k^\dagger q_c) 
    (q_c^T I_\alpha(|\omega +k \omega_d |) q_a)(q_b^T I_\beta(\omega) q_c)}
    {(\Omega_a^2 - (\omega+k \omega_d  - i\Gamma_a)^2)
    (\Omega_b^2 - (\omega - i\Gamma_b)^2)
    (\Omega_c^2 - (\omega +k \omega_d  + i\Gamma_c)^2)
    (\Omega_d^2 - (\omega + i\Gamma_d)^2)}
\end{equation}
In the weak coupling limit where $\Gamma_a \ll
\Omega_a, \omega_d$ and under the condition
$\omega_d < \min_{\Omega_a \neq \Omega_b}\{|\Omega_a - \Omega_b|/2\}$ 
the typical shape of the functions
$p_{\alpha,\beta}^{(k)}(\omega)$ is like the one depicted 
in figure \ref{fig:example_wc}.
We see the expected resonances at frequencies $\{\Omega_a\}$ 
and $\{\Omega_a - k\omega_d\}$, which for the mentioned conditions are 
well defined and do not overlay.
Therefore, an approximate solution for integrals of the form $\int_0^\infty 
p_{\alpha,\beta}^{(k)}(\omega) N(\omega)$ can be obtained by
dealing separately with each resonance peak, evaluating the remaining
factors at the center of the peak. It is clear that the dominant terms 
are those with $\Omega_c = \Omega_a$ and $\Omega_d = \Omega_b$. Also, for low
temperatures the Planck distribution $N(\omega)$ exponentially supresses high
frequencies, and therefore the dominant contribution to the integral is given
by the peak at $\omega = \Omega_0 -k\omega_d$, where $\Omega_0$ is the smallest
normal frequency of the system. Using this method the following expression
is found:
\begin{equation}
    \int_0^\infty d\omega \; p_{\alpha,\beta}^{(k)}(\omega) N(\omega) \simeq
    \frac{\pi^2}{8} \frac{N(\Omega_0-k\omega_d)}{\Omega_0^2\Gamma_0}
    \sum_{b,c} \frac{(q_0^tV_kq_b)(q_b^tV_k^\dagger q_c) (q_b^t
    I_\beta(\Omega_a-k\omega_d)q_c) (q_0^tI_\alpha(\Omega_a)q_0)}
    {(\Omega_b^2 -(\Omega_0 -k\omega_d)^2)
    (\Omega_c^2 -(\Omega_0 -k\omega_d)^2)}
    \label{eq:int_p_N}
\end{equation}
Note that this expression is invariant upon time inversion of the process
(i.e, invariant under complex conjugation of the Fourier coefficients $V_k$).
We now use the previous result to calculate the resonant contributions to the
heat rates for the simple driving protocol $V(t)=V_0+2V_1\cos(\omega_d t)$
in a system connected to only two reservoirs, $E_\alpha$ and $E_\beta$, which
are at the same temperature $T_0$.
Considering, for simplicity, only the contribution of the $\Omega_0$ normal
mode, we have:
\begin{equation}
\begin{split}
    \dot {\bar Q}_\alpha^{R} = \dot {\bar Q}_\alpha^{RP} + \dot {\bar Q}_\alpha^{RH} &= 
    \frac{\pi^2}{8} \frac{N(\Omega_0-\omega_d)}{\Omega_0^2\Gamma_0}
    \frac{|V_1^0|^2}{(\Omega_0^2 -(\Omega_0 -\omega_d)^2)^2}\\
    &\times \left\{ 
    (\Omega_0-\omega_d) I^0_\beta(\Omega_0) I^0_\alpha(\Omega_0-\omega_d)
    -\Omega_0 I^0_\alpha(\Omega_0)I^0_\beta(\Omega_0-\omega_d)
    -k \omega_d I^0_\alpha(\Omega_0) I^0_\alpha(\Omega_0-\omega_d)
    \right\},
    \label{eq:wc_QRP}
\end{split}
\end{equation}
where we have defined $M^0=q_0^t M q_0$ for any matrix $M$. 
It is easy to see that if the two reservoirs are spectrally equivalent
(if $I_\alpha(\omega)=I_\beta(\omega)$), then the previous expression is always
negative and therefore both reservoirs are heated. However, if the spectral
densities satisfy $I_\alpha(\Omega_0) \ll I_\beta(\Omega_0)$ then the last two
terms between brackets in the previous expression can be neglected with
respect to the first, and $\dot{\bar Q}^R$ becomes positive. Therefore, the
condition $I_\alpha(\Omega_0) \ll I_\beta(\Omega_0)$ allows cooling of the
reservoir $E_\alpha$.

\subsection{Minimum temperature}

For low temperatures the Planck distribution in Eq. (\ref{eq:wc_QRP}) can be
approximated by $N(\Omega_0 - \omega_d) \simeq e^{-(\Omega_0-\omega_d)/T_0}$,
which vanishes faster than any power law as $T_0 \to 0$. This strong dependence
for low temperatures makes it imposible to reach zero temperature in finite
time. However, as explained in the main text, this effect can be avoided by
instantaneously adjusting the driving frequency $\omega_d$ as $T_0$ decreases, 
in such a way that $\Omega_0-\omega_d \simeq T_0$. If we assume that
$I_\alpha(\omega) \propto \gamma_0 \omega^{\lambda_\alpha}$ for low
frequencies, then it is clear that the resonant heat rate in Eq.
(\ref{eq:wc_QRP}) scales as $\dot {\bar Q}_\alpha^R \propto \gamma_0
(\Omega_0-\omega_d)^{1+\lambda_\alpha}$ (the factor
$I_\beta(\Omega_0)/\Gamma_0$ is independent of the coupling constant between
the system and reservoir $E_\beta$). 
Thus, for the mentioned adaptative strategy, we have 
$\dot {\bar Q}_\alpha^\text{R} \propto \gamma_0 T_0^{1+\lambda_\alpha}$.
On the other hand, it can be seen from Eq. (\ref{eq:heat_aux_1}) that for $\omega_d < \Omega_0$
the non resonant heating $\dot{\bar Q}_\alpha^{\text{NRH}}$ is proportional to
$\gamma_0^2$ (since the integration domain does not include any resonance peak
of the function $p_{\alpha,\beta}^{-1}(\omega)$). Also, for low driving
frequency $\omega_d$ it scales as $\omega_d^{2+2\lambda_\alpha}$. Thus, for the
adaptative strategy, we obtain $\dot{\bar Q}_\alpha^{\text{NRH}} \propto
\gamma_0^2 (\Omega_0-T_0)^{2+2\lambda_\alpha}$. 

Therefore, we see that there always exists a temperature $T_{min}$ below which
$|\dot{\bar Q}_\alpha^{\text{NRH}}| > \dot {\bar Q}_\alpha^\text{R}$ and
the net effect is to heat up the reservoir $E_\alpha$. Also, 
this minimum temperature scales as $T_{min} \propto
\gamma_0^{1/(1+\lambda_\alpha)}$ (for $T_{min} \ll \Omega_0)$. 

\subsection{Numerical evaluation of the heat rates for a simple case}
\begin{figure}
  \centering
  \begin{minipage}[b]{0.4\textwidth}
    \includegraphics[width=\textwidth]{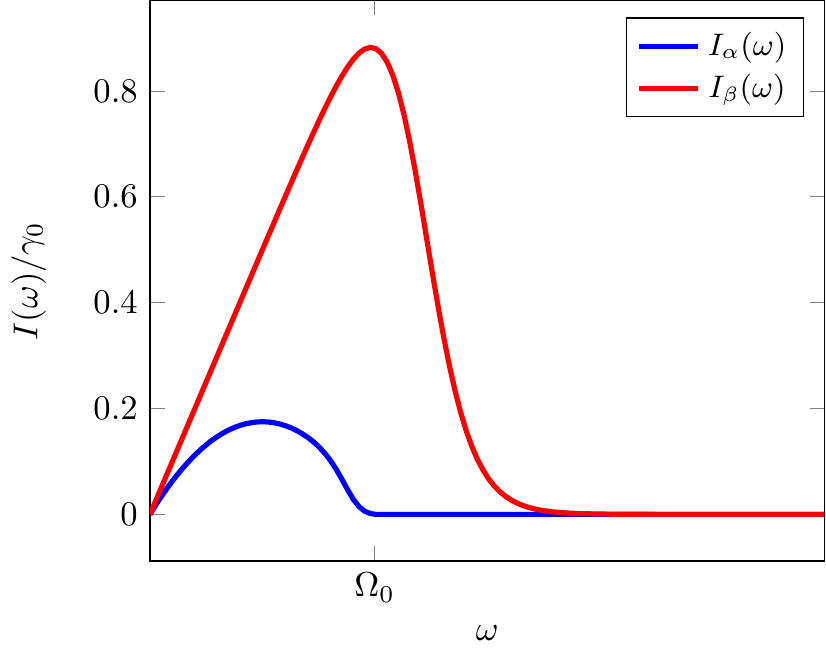}
    \caption{\label{fig:spectral} Spectral densities used for
the numerical evalution of the heat rates
($\lambda_\alpha\!=\!\lambda_\beta\!=\!1$). 
They exactly satisfy the cooling
condition $I_\alpha(\Omega_0)\!\ll\! I_\beta(\Omega_0)$.}
  \end{minipage}
  \hspace{1cm}
  \begin{minipage}[b]{0.4\textwidth}
    \includegraphics[width=\textwidth]{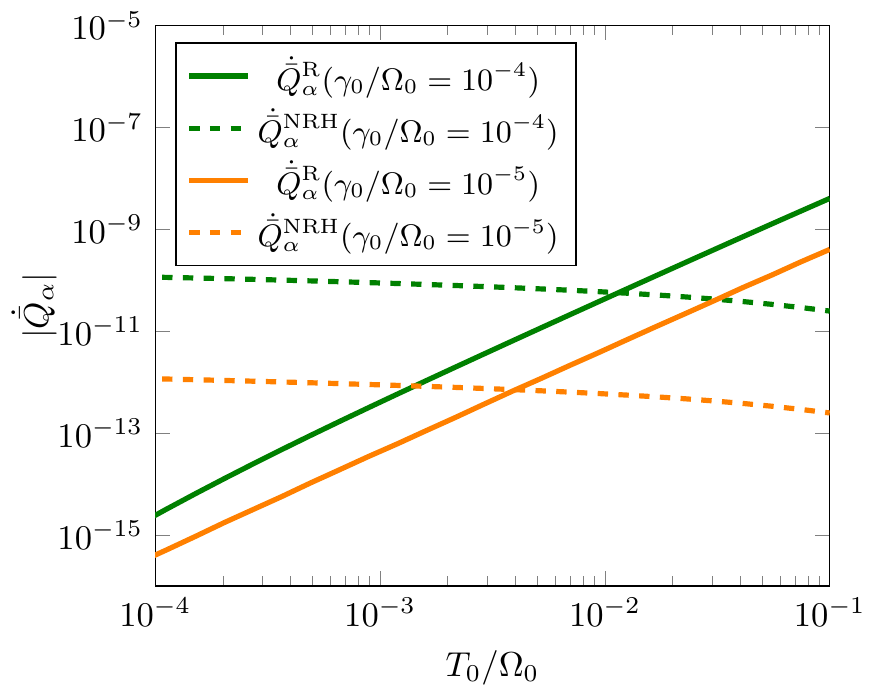}
    \caption{\label{fig:example_heat} Resonant and non-resonant contributions
to the total heat rate $\dot{\bar Q}_\alpha$. The driving frequency is
$\omega_d = \Omega_0 - T_0$ and $\lambda_\alpha=1$.}
  \end{minipage}
\end{figure}

In order to test the last result we evaluated numerically the heat rates in a particular case.
We consider a system composed of a single harmonic oscillator of bare frequency
$\Omega_0$ coupled to two reservoirs, $E_\alpha$ and $E_\beta$, with spectral
densities given by:
\begin{equation}
\begin{split}
    I_\alpha(\omega) &= \gamma_\alpha \; \omega^{\lambda_\alpha} \; (\Omega_0 - \omega) \;
    \theta((\omega-\Lambda_\alpha)/r_\alpha) \qquad \qquad (\omega < \Omega_0)\\
    I_\beta(\omega) &= \gamma_\beta \; \omega^{\lambda_\beta} \; \theta((\omega-\Lambda_\beta)/r_\beta)\\
    \label{eq:spectral}
\end{split}
\end{equation}
where $\theta(x) = e^{-x}/(1+e^{-x})$ is an exponential cutoff. Note that 
$I_\alpha(\omega)$ vanishes at $\omega=\Omega_0$. In this way, the cooling
condition $I_\alpha(\Omega_0) \ll I_\beta(\Omega_0)$ is exact.
In particular, we choose the parameters $\gamma_\beta=\gamma_0$,
$\gamma_\alpha=0.7\gamma_0$ (where $\gamma_0$ 
is a coupling constant that will be varied), $\Lambda_\alpha=0.9$,
$\Lambda_\beta=1.2$,$r_\alpha=0.04$, $r_\beta=0.1$ (all this parameters are in
units of $\Omega_0$). In figure \ref{fig:spectral} we show the spectral
densities for the case in which both are ohmic ($\lambda_\alpha = \lambda_\beta
= 1$).

For these spectral densities, the contributions $\dot{\bar Q}_\alpha^{\text{NRH}}$
and $\dot{\bar Q}_\alpha^{\text{R}} = \dot{\bar Q}_\alpha^{\text{RP}} + \dot{\bar Q}_\alpha^{\text{RH}}$
to the heat rates are calculated by numerical integration of the expressions
given in Eqs. (\ref{meq:heat_rp}), (\ref{meq:heat_rh}), and
(\ref{meq:heat_nrh}) in the main text (under the weak driving approximation).
As an example, we plot in Figure \ref{fig:example_heat} these resonant and
non-resonant contributions (in absolute value) versus the common temperature
$T_0$, for two different values of the
coupling constant $\gamma_0$. These results corresponds to the adaptative
strategy for which the driving frequency is selected as $\omega_d = \Omega_0 -
T_0$. We see that $\dot{\bar Q}_\alpha^{\text{NRH}}$
scales as $\gamma_0^2$ while $\dot{\bar Q}_\alpha^{\text{R}}$ scales as
$\gamma_0$. The temperature for which $\dot{\bar Q}_\alpha^{\text{NRH}}$ and 
$\dot{\bar Q}_\alpha^{\text{R}}$ become equal is the minimum temperature
$T_{min}$ for which the adaptative strategy supports cooling of reservoir
$E_\alpha$. The dependence of $T_{min}$ with the coupling constant $\gamma_0$ is
shown in Figure \ref{fig:min_temp} in the main text for $\lambda_\alpha=1$ and
$\lambda_\alpha=2$ and is found to be well described by the power law discussed
above.

\end{document}